\documentclass[12pt,a4paper,reqno]{article}

\usepackage{latexsym}
\usepackage[english]{babel}
\usepackage{fancyhdr}
\usepackage[mathscr]{eucal}
\usepackage{amsmath}
\usepackage{mathrsfs}
\usepackage{amsthm}
\usepackage{amsfonts}

\usepackage{amssymb}
\usepackage{amscd}
\usepackage{graphicx}
\usepackage{graphics}
\usepackage{latexsym}
\usepackage{color}

\theoremstyle{plain}
\numberwithin{equation}{section}

\newtheorem{theorem}{Theorem}[section]

\newtheorem{lemma}[theorem]{Lemma}
\newtheorem{proposition}[theorem]{Proposition}
\newtheorem{definition}[theorem]{Definition}

\newtheorem{assumption}[theorem]{Assumption}
\newtheorem{remark}[theorem]{Remark}

\newcommand {\absleq} {{\leq_{|\, \cdot\, |}\, }}

\def\tD{{\tilde{\mathcal D}}}

\def\ben{\begin{enumerate}}
\def\een{\end{enumerate}}
\def\bgdf{\begin{definition}}
\def\eddf{\end{definition}}
\def\bgass{\begin{assumption}}
\def\edass{\end{assumption}}
\def\bglm{\begin{lemma}}
\def\edlm{\end{lemma}}
\def\bgpf{\begin{proof}}
\def\edpf{\end{proof}}
\def\bgth{\begin{theorem}}
\def\edth{\end{theorem}}
\def\bgcor{\begin{collorary}}
\def\edcor{\end{collorary}}
\def\bgprp{\begin{proposition}}
\def\edprp{\end{proposition}}
\def\bgrm{\begin{remark}}
\def\edrm{\end{remark}}

\def\lbeq(#1){\label{eqn:#1}}
\def\refeq(#1){{\rm (\ref{eqn:#1})}}
\def\refeqs(#1,#2){{\rm (\ref{eqn:#1}) and (\ref{eqn:#2})}}
\def\refeqss(#1,#2,#3){{\rm (\ref{eqn:#1}),\, (\ref{eqn:#2}) and (\ref{eqn:#3})}}
\def\lbth(#1){\label{th:#1}}
\def\refth(#1){{\rm Theorem \ref{th:#1}}} 
\def\refths(#1,#2){{\rm Theorems \ref{th:#1} or \ref{th:#2}}}
\def\refthss(#1,#2,#3){{\rm Theorems \ref{th:#1}, \ref{th:#2} or \ref{th:#3}}}
\def\refthb(#1){{\bf Theorem \ref{th:#1}}}
\def\lblm(#1){\label{lm:#1}}
\def\reflm(#1){{\rm Lemma \ref{lm:#1}}}
\def\reflms(#1,#2){{\rm Lemmas \ref{lm:#1} and \ref{lm:#2}}}
\def\reflmss(#1,#2,#3){{\rm Lemmas \ref{lm:#1}, \ref{lm:#2} and \ref{lm:#3}}}
\def\reflmsss(#1,#2,#3,#4){{\rm Lemmas \ref{lm:#1},\, \ref{lm:#2},\, \ref{lm:#3} and \ref{lm:#4}}}
\def\reflmb(#1){{\bf Lemma \ref{lm:#1}}}
\def\lbprp(#1){\label{prp:#1}}
\def\refprp(#1){{\rm Proposition \ref{prp:#1}}}
\def\refprpb(#1){{\bf Proposition \ref{prp:#1}}}
\def\lbcor(#1){\label{cor:#1}}
\def\refcor(#1){{\rm Corollary \ref{cor:#1}}}
\def\refcors(#1,#2){{\rm Corollaries \ref{cor:#1} and \ref{cor:#2}}}
\def\lbrm(#1){\label{rm:#1}}
\def\refrm(#1){{\rm Remark \ref{rm:#1}}}
\def\lbass(#1){\label{ass:#1}}
\def\refass(#1){{\rm Assumption \ref{ass:#1}}}
\def\refasss(#1,#2){{\rm Assumption \ref{ass:#1} and \ref{ass:#2}}}
\def\lbdf(#1){\label{df:#1}}
\def\refdf(#1){{\rm Definition \ref{df:#1}}}
\def\lbsec(#1){\label{s:#1}}
\def\refsec(#1){{\rm \S\ref{s:#1}}}
\def\lbsubsec(#1){\label{ss:#1}}
\def\refsubsec(#1){{\rm \S\ref{ss:#1}}}

\def\tGa{\widetilde{\Gamma}}
\def\Bg{{\mathcal B}}
\def\Cg{{\mathcal C}}
\def\Fg{{\mathcal F}}
\def\Gg{{\mathcal G}}

\def\Vg{{\mathcal V}}

\newcommand{\lam}{\lambda}
\def\ab{{\bf a}}
\def\bb{{\bf b}}

\def\Bb{{\bf B}}

\def\eb{{\bf e}}
\def\fb{{\bf f}}

\def\yb{{\bf y}}
\def\ph{{\varphi}}
\def\bqn{\begin{equation}}
\def\eqn{\end{equation}}
\def\C{{\mathbb C}}

 \def\Cb{{\overline{\mathbb C}}}

\def\R{{\mathbb R}}
\def\Rg {{\mathcal R}}
\def\a{\alpha}

\def\c{\gamma}
\def\Ga{\Gamma}
\def\d{\delta}
\def\Dg{{\mathcal D}}
\def\Eg{{\mathcal E}}
\def\Mg{{\mathcal M}}
\def\Sg{{\mathcal S}}

\def\ep{\varepsilon}

\def\th{\theta}

\def\k{\kappa}
\def\m{\mu}
\def\n{\nu}
\def\r{\rho}
\def\s{\sigma}
\def\t{\tau}
\def\w{\omega}
\def\W{\Omega}
\def\Hg{{\mathcal H}}
\def\la{\langle}
\def\ra{\rangle}

\def\lap{\Delta}
\def\ax{{\la x \ra}}
\def\pa{{\partial}}

\def\br{\begin{array}}
\def\er{\end{array}}
\def\Ker{{\rm Ker\,} }
\def\ax{{\la x \ra}}
\def\ay{{\la y \ra}}

\begin{document}

\title
{{$L^p$-boundedness of wave operators for 2D 
Schr\"odinger operators with point interactions}}

\author{Kenji Yajima \footnote{Department of Mathematics, Gakushuin University, 
1-5-1 Mejiro Toshima-ku Tokyo 171-8588 (Japan), Supported by JSPS 
grant in aid for scientific research No. 19K03589}}

\date{}

\allowdisplaybreaks
\maketitle

\vspace{-0.5cm}
\begin{center}
{\it Dedicated to Professor Arne Jensen on the occasion of 
  his 70th birthday}
\end{center}

\vspace{0.2cm}

\begin{abstract}
For two dimensional Schr\"odinger operator $H$ with point interactions,  
we prove that wave operators of scattering for the pair $(H,H_0)$, 
$H_0$ being the free Schr\"odinger operator, are bounded in the Lebesgue space 
$L^p(\R^2)$ for $1<p<\infty$ if and only if there are no generalized 
eigenfunctions of $Hu(x)=0$ which satisfy $u(x)= C|x|^{-1}+ o(|x|^{-1})$ 
as $|x|\to \infty$, $C\not=0$. Otherwise they are bounded for $1<p\leq 2$ 
and unbounded for $2<p<\infty$.  
\end{abstract}

\section{Introduction} 

We consider Schr\"odinger operators in $\Hg= L^2(\R^2)$ with point interactions 
(SOPI in short) at $Y=\{y_1, \dots, y_N\}\subset \R^2 $ with strength 
$\a= (\a_1, \dots, \a_N)\in \R^N$, $1\leq N <\infty$ which are defined 
symbolically by 
\bqn \lbeq(1-1)
H_{\a,Y}= ``-\lap + \sum_{j=1}^N \a_j \d(x-Y_j)''
\eqn 
and which will shortly be defined rigorously. Solutions of $H_{\a,Y}u=0$ 
which are bounded as $|x|\to 0$ are called (threshold) resonance. We  show 
that a resonance satisfies $u(x) = a+ b\cdot{x}/|x|^2+ O(|x|^{-2})$ 
as $|x|\to \infty$ for a constant $a\in \C $ and a vector $b\in \C^2$; 
we call it $s$-wave 
resonance if $a\not=0$, $p$-wave resonance if $a=0$ and $b\not=0$; it is a 
zero energy eigenfunction if $a$ and $b$ vanish but $u\not=0$. It wil be   
shown which kind of resonances $H_{\a,Y}$ can possess is controled 
by three $N \times N$ symmetric matrices defined in terms of $\a$ and $Y$.
We then prove that the wave operators of scattering for the pair 
$(H_{\a,Y}, H_0)$, $H_0=-\lap$ being the free Schr\"odinger operator, 
are bounded in the Lebesgue space $L^p(\R^2)$ for all $1<p<\infty$ 
if and only if $p$-wave resonances are absent from $H_{\a,Y}$ and, 
otherwise they are bounded for $1<p \leq 2$ and unbounded for $2<p<\infty$. 

For the roles played by SOPI in physics, in nuclear and solid state physics in 
particular, and for the history of its mathematical studies, 
we refer to the seminal monograph \cite{AGHH}, the introduction of 
\cite{CMY} and references therein and we start with reviewing the rigorous 
definition of $H_{\a,Y}$ and some of its basic properties (\cite{AGHH}). 
The resolvent  $G_0(z)= (H_0-z^2)^{-1}$ of the free Schr\"odinger operator 
with the momentum parameter 
$z \in \C^{+}=\{z\in \C \colon \Im z >0\}$, 
is the convolution operator with 
\begin{equation}\lbeq(Gg)
\mathcal{G}_{z}(x)
\stackrel{\rm def}{=}
\frac1{(2\pi)^2}\int_{\R^2}\frac{e^{ix\xi}d\xi}{\xi^2-z^2}
= \frac14 H_0^{(1)}(z|x|), 
\end{equation}
where $H_0^{(1)}(z)$ is the Hankel function of the first kind:
\begin{align}
& \frac14 H_0^{(1)}(z) = \left(-\frac1{2\pi}\log\Big(\frac{z}{2i}\Big)-
\frac{\gamma}{2\pi}\right) \sum_{k=0}^\infty \frac{(-1)^k}{(k!)^2}
\left(\frac{z^2}{4}\right)^k \notag \\
&\ \  -\frac1{2\pi}\left(
\frac{\frac{1}{4}z^2}{(1!)^2}
-\left(1+\frac12\right)\frac{\left(\frac14{z^2}\right)^2}{(2!)^2}
+\left(1+\frac12+\frac13\right)\frac{\left(\frac14{z^2}\right)^3}{(3!)^2}- 
\cdots \right)  \lbeq(hankel-1) \\
& \quad = \frac{e^{iz}}{2^\frac32 \pi}\int_0^\infty e^{-t}t^{-\frac12}
\left(\frac{t}2 - iz\right)^{-\frac12}dt, \quad 
z\in \Cb^{+}\setminus \{0\}, \lbeq(hankel-2)
\end{align}
where $\gamma$ is Euler's constant (\cite{Watson}). 
We denote the prefactor in \refeq(hankel-1) by $g(z)$:    
\bqn 
g(z)= -\frac1{2\pi}\log\Big(\frac{z}{2}\Big)+ \frac{i}{4}- 
\frac{\gamma}{2\pi}, \lbeq(g-def)
\eqn 
where $\log(z/2)$ is real for $z>0$. 
Notice that with two dimensional Newton potential $N_0(x)$ 
\bqn 
g(z|x-y|)= g(z) + N_0(x-y), \quad N_0(x)= -(2\pi)^{-1}\log |x| 
\eqn 
is the leading term of the expansion of $\Gg_z(x-y)$ as $z \to 0$. 
Define $N \times N$ matrix $\Ga_{\a,Y}(z)$ for 
$z\in \Cb^{+}=\{z\in \C \colon \Im z\geq 0\}$ by 
\bqn \lbeq(Ga-def) 
\Ga_{\a,Y}(z) = 
\left\{(\a_j -g(z))\d_{jk} - \Gg_z(y_j-y_k)\hat{\d}_{jk}\right\},
\eqn  
where $\d_{jk}$ is the Kronecker delta and $\hat{\d}_{jk}= 1 -\d_{jk}$. 
It is shown (cf. \cite{AGHH}) that $\Ga_{\a,Y}(z)$, $z\in \C^{+}$ 
is non-singular outside a finite subset $\Eg \subset i(0,\infty)$ 
and the operator valued 
function $R(z^2)$ defined for $z \in \C^{+}\setminus \Eg$ by 
\bqn  \lbeq(HaY-def)
R(z^2)=(H_0-z^2)^{-1} + \sum_{j,k=1}^N 
[\Ga_{\a, Y}(z)^{-1}]_{jk} \Gg_z(\cdot - y_j) \otimes 
\overline{\Gg_z(\cdot-y_k)}
\eqn 
is the resolvent of the seladjoint operator $H_{\a,Y}$ in $\Hg$: 
$R(z^2)=(H_{\a,Y}-z^2)^{-1}$; $H_{\a,Y}$ is the 
selfadjoint extension of $-\lap\vert_{C_0^\infty(\R^2\setminus Y)}$ 
formally defined by \refeq(1-1);
it is a real local operator; domain $D(H_{\a,Y})$ is the set of $u$'s 
of the form 
\bqn \lbeq(DOMH)
u(x)= v(x)+ \sum_{j,k=1}^N 
[\Ga_{\a, Y}(z)^{-1}]_{jk} v(y_k)\Gg_z(x-y_j), \ v \in H^2(\R^2);
\eqn   
the function $u$ determines $v$ uniquely in \refeq(DOMH) and 
$(H_{\a,Y}-z^2) u= (H_0 - z^2)v$,  $H^2(\R^2)$ being the Sobolev space 
of secon order. 

Spectrum of $H_{\a,Y}$ consists of the 
absolutely continuous (AC for short) part $[0,\infty)$ and at most 
$N$ number of non-positive eigenvalues. 
The definition \refeq(HaY-def) shows that the rank of 
$R(z^2)-(H_0-z^2)^{-1}$ is $N$ and, Kato-Rosenblum theorem (\cite{Kato, RS}) 
implies that the wave operators 
defined by the strong limits in $L^2(\R^2)$:
\bqn \lbeq(limit)
W_{\a,Y}^{\pm}= \lim_{{ t\to\pm \infty}} e^{itH_{\a,Y}}e^{-itH_0} 
\eqn 
exist and are complete in the sense that 
${\rm Range}\ W_\pm =L^2_{ac}(H_{\a,Y})$, 
the AC subspace of $L^2(\R^2)$ for $H_{\a,Y}$. 
In this paper we study if the wave operators $W_{\a,Y}^{\pm}$ 
are bounded in  $L^p(\R^2)$, $1\leq p\leq \infty$. 

We introduce the three real symmetric matrices which will play important roles 
in the rest of the paper: 
\begin{gather}
\lbeq(deftD)
\tD= \Big(\d_{jk}\a_j + \frac{\hat{\d}_{jk}}{2\pi}\log|y_j-y_k|\Big), 
\ \ 
\Gg_1(Y)= - \Big(\frac{\hat\d_{jk}}{4N}|y_j-y_k|^2\Big), \\ 
\Gg_2(Y)= - \left(\frac{\hat\d_{jk}}{8{\pi}N}|y_j-y_k|^2 
\log\Big(\frac{e}{|y_j-y_k|}\Big) \right)\, 
\end{gather}
and, which appear in the asymptotic expansion as $\lam \to 0$ of $\Ga(\lam)$: 
\bqn \lbeq(expansion-1)
\Ga(\lam) = -Ng\Big(P - \frac{g^{-1}\tD}{N} + \lam^2 \Gg_1(Y) + 
\lam^2 g(\lam)^{-1}
\Gg_2(Y) + O(\lam^4) \Big). 
\eqn 
Here $P$ and $S$ are projections in $\C^N$:
\begin{align}\lbeq(horia1)
\eb =  
\frac{1}{\sqrt{N}}{{\bf 1}}, \quad  
{\bf 1}= \begin{pmatrix} 1\\  \vdots \\ 1 
\end{pmatrix}, \quad 
P = \eb \otimes \eb, \quad S=1-P.
\end{align}
It is known (\cite{CMY}) that these matrices control also the 
the asymptotic behavior as $z \to 0$ of $(H_{\a,Y}-z^2)^{-1}$ and 
threshold resonances which $H_{\a,Y}$ can have. 
We shall make the latter point clear by defining the resonanaces 
as zero energy solutions $H_{\a,Y} \ph =0$ in an weighted $L^2$ spaces.

The following is the main theorem of this paper. It will be stated 
by using the matrices defined above and it appears sightly differently 
from what is stated at the beginning of the paper, however, it will 
shortly become clear that they are actually equivalent. 
\bgth \lbth(Main) \ben
\item[{\rm (1)}] Suppose that linear map $S{\tD} S$ in $S\C^N$ is 
non-singular. Then  $W_{\a,Y}^{\pm}$ are bounded from $L^p(\R^2)$ to itself 
for all $1<p<\infty$. 
\item[{\rm (2)}] Suppose $S{\tD} S$ is singular in $S\C^N$ and let 
$T$ be orthogonal projection in $S\C^n$ onto ${\rm Ker}_{S\C^N}\, S{\tD}S$.
Suppose $T\tD^2 T$ is non-singular in $T\C^N$. 
Then  $W_{\a,Y}^{\pm}$ are bounded from $L^p(\R^2)$ to itself 
for all $1<p<\infty$. 
\item[{\rm (3)}] Suppose $S{\tD} S$ is singular in $S\C^N$ and that 
$T\tD^2 T$ is also singular in $T\C^N$. Let $T_p$ 
be orthogonal projection in $T\C^n$ onto ${\rm Ker}_{T\C^N}\, T{\tD}^2 T$. 
Suppose $T_p \Gg_1(Y)T_p$ is non-singular in $T_p\C^N$. 
Then, $W_{\a,Y}^{\pm}$ are bounded from $L^p(\R^2)$ to itself 
for $1<p\leq 2$ but are unbounded for $2<p<\infty$. 
\item[{\rm (4)}] Suppose $S{\tD} S$ is singular in $S\C^N$, 
$T\tD^2 T$ in $T\C^N$ and that $T_p \Gg_1(Y)T_p$ is also singular 
in $T_p\C^N$. Let $T_e$ be the orthogonal projection in $T_p\C^N$ onto 
$\Ker_{T\C^N} T_p \Gg_1(Y)T_p$. Suppose in addition 
$T_e\not=T_p$, viz. 
$T_p \Gg_1(Y)T_p\not=0$. Then, 
$W_{\a,Y}^{\pm}$ are bounded from $L^p(\R^2)$ to itself 
for $1<p\leq 2$ and unbounded for $2<p<\infty$. 
\item[{\rm (5)}] Suppose $S{\tD} S$ is singular in $S\C^N$, 
$T\tD^2 T$ in $T\C^N$ and $T_p \Gg_1(Y)T_p=0$. 
Then, $W_{\a,Y}^{\pm}$ are bounded from 
$L^p(\R^2)$ to itself for all $1<p<\infty$.
\een
\edth 

\bgrm 
\ben 
\item[{\rm (1)}] Statements {\rm (3)} and  {\rm (4)} may of course be 
unified simply by assuming 
$T_p \Gg_1(Y)T_p\not=0$. We state \refth(Main) in this way 
only for a later convenince.
\item[{\rm (2)}]  
It is known {\rm (\cite{CMY})} that 
${\rm rank}\, T\tD^2 T \leq 1$ under the condition of statement {\rm (2)} 
and, in statements {\rm (4)} and {\rm (5)}, $T_e \Gg_2(Y)T_e$ is necessarily 
non-singular in $T_e \C^N$.  
\een
\edrm

For regular Schr\"odinger operators $H= -\lap + V(x)$ on $\R^d$, 
$L^p$-boundedness of wave operators has long been studied  
and many results are known under various assumptions on $V$. 
Results depend on the dimensions $d$ and on the existence/absence of 
eigenvalue and/or resonances at $z=0$. We list here some of the results.   
In the following it is assumed that $|V(x)|\leq C \ax^{-\s}$ for  
$\s>2$ or for a larger $\s$.  
\ben 
\item[{\rm (1)}] If $d=1$, $W^\pm$ are bounded in $L^p(\R)$ for $1<p<\infty$ 
but not for $p=1$ and $p=\infty$ (\cite{RW,GY, DF}).   
\item[{\rm (2)}] If $H$ has no eigenvalue nor resonances at $z=0$,  
$W^\pm$ are bounded in $L^p(\R^d)$ for $1\leq p\leq \infty$ if $d \geq 3$ 
and for $1<p<\infty$ if $d=2$ (\cite{Y-3dim, Y-2dim, JY-2}).   
\item[{\rm (3)}]  If $H$ has an eigenvalue or resonances at $z=0$,  
much is known if $d\geq 5$ or $d=3$ and the results depend on $d$ and 
the types of sigularities of the resolvent at $z=0$ 
(\cite{KY,Y-3d-sing, FY, EG, GG, EGG}).  
\item[{\rm (4)}] If $d=4$ and $H$ has an eigenvalue but no resonances at $0$, 
$W^\pm$ are bounded in $L^p(\R^4)$ for $1\leq p \leq 4$ (\cite{JY-4, GG}).
\item[{\rm (5)}] If $d=2$ and $H$ has an $s$ wave resonances or only eigenvalue 
at $0$, $W^\pm$ are bounded in $L^p(\R^2)$ for $1<p <\infty$ (\cite{EGG}).
\een
For SOPI, $W_{\a, Y}^\pm $ are bounded in 
$L^p(\R)$ for $1<p <\infty$ for all $\a$ and $Y$ if $d=1$ (\cite{DMW}); 
if $H_{\a, Y}$ has no eigenvalue nor resonances 
at zero, then $W_{\a, Y}^\pm $ are bounded in $L^p(\R^2)$ for $1<p <\infty$ 
if $d=2$ (\cite{CMY}) and in $L^p(\R^3)$ for $1<p<3$ if $d=3$ (\cite{DMSY}). 
Thus, \refth(Main) gives a complete result for SOPI in two dimensions,  
however, the problem for the end points $p=1$ and $p=\infty$ are still open. 
We mention that for Schr\"odinger operators with regular potentials 
in two dimensions, no results have been obtained when $H$ has $p$-wave resonances 
which corresponds to the case of statements (3) and (4) of \refth(Main). 

The three matrices $\tD, \Gg_1(Y)$ and $\Gg_2(Y)$ control 
threshold resonances and the asymptotic behavior of the resolvent 
$(H_{\a,Y}-z^2)^{-1}$ as $z \to 0$. We introduce some notation. 
For $\s\in \R$, $L_{\s}^2$ and $H^2_{\s}$ are weighted spaces:  
\begin{gather*}
L^2_{\s}(\R^2)\stackrel{\rm def}{=} 
 \{\ax^{-\s}u(x) \colon u \in L^2(\R^2)\}, 
\quad \|u\|_{L^2_{\s}}\stackrel{\rm def}{=}  \|\ax^{\s}u\|_{L^2}, \\
H^2_{\s}(\R^2) \stackrel{\rm def}{=} 
\{\ax^{-\s} u \colon u \in H^2(\R^2)\}, \quad 
\|u\|_{H^2_\s}\stackrel{\rm def}{=}  \|\ax^{\s} u\|_{H^2}.
\end{gather*}
For $y\in \R^2$, $\t_{y} u(x) = u(x-y)$ is the translation by $y$ 
and we set 
\[
\hat{v}_Y (x)= 
\begin{pmatrix} \t_{y_1}v(x) \\ \vdots \\ \t_{y_N}v(x) \end{pmatrix}, \ \   
\hat{\Gg}_{z,Y}(x)= \begin{pmatrix} \t_{y_1}\Gg_z(x) \\ \vdots 
\\ \t_{y_N}\Gg_z(x) \end{pmatrix}, \ \ 
\hat{N}_{0,Y} (x)= \begin{pmatrix} \t_{y_1}N_0(x) \\ \vdots \\ \t_{y_N}N_0(x). 
\end{pmatrix}
\]
In terms of these vectors, domain of $H_{\a,Y}$ is given by 
\bqn \lbeq(domainof)
D(H_{\a,Y})= 
\{u(x)= v(x)+ \la  
\Ga_{\a, Y}(z)^{-1} v_Y, {\Gg}_{z,Y}(x)\ra  \colon v \in H^2(\R^2)\}.
\eqn  
Here and hereafter $\la \ab, \bb \ra = a_1 b_1 + \cdots + a_N b_N$ 
without complex conjugation. We shall often write 
\[
a \absleq b \ \ \mbox{for} \ \ |a|\leq |b|.
\]

In view of the proof of the corresponding statement for $H_{\a,Y}$ 
in \cite{AGHH} the following lemma should be obvious and the proof 
will be omitted. 
\bglm \lblm(1-1) Let $1<\s<2$ and $z\in \C^{+}\setminus \Eg$. 
The operator $R(z^2)$ defined by {\rm \refeq(HaY-def)} 
can be extended to a bounded operator 
in $L^2_{-\s}(\R^2)$ by continuity, which we denote by $R_{-\s}(z^2)$. 
Then, $R_{-\s}(z^2)$ is the resolvent of 
a closed operator $H_{\a,Y}^{-\s}$ 
in $L^2_{-\s}(\R^2)$.  Domain of  $H_{\a,Y}^{-\s}$ is given 
independently of $z\in \C^{+}\setminus \Eg$ by 
\bqn \lbeq(dom-h)
{\rm Image}\, R_{-\s}(z^2)
=\{u= v+ \la  
\Ga_{\a, Y}(z)^{-1} v_Y, \hat{\Gg}_{z,Y}\ra  \colon v \in H_{-\s}^2(\R^2)\},
\eqn 
where $v\in H_{-\s}^2(\R^2)$ is uniquely determined by $u$  and 
\bqn \lbeq(def-of-hweight)
(H_{\a,Y}^{-\s}-z^2) u = (H_0-z^2)v.
\eqn 
\edlm 

\bglm\lblm(1-2)  The null space of $H_{\a, Y}^{-\s}$ is given 
independently of $1<\s<2$ by  
\bqn \lbeq(phlog) 
\Ker H_{\a, Y}^{-\s} =\Big\{\ph(x) 
= \frac{\la\tD\ab, {{\bf 1}} \ra}{N}  -\frac1{2\pi}  
\sum_{j=1}^N a_j \log |x-y_j| \colon \ab\in \Ker S\tD\Big\},
\eqn 
where $a_1, \dots, a_N$ are components of $\ab \in \C^N$. 
\edlm 

We denote by $C_b(\R^2 \setminus Y)$ the set of continous functions in 
$\R^2 \setminus Y$ which are bounded outside a bounded open set containing 
$Y$ and define 
$\Rg_{\a,Y}=\Ker H^{-\s}_{\a,Y} \cap C_b(\R^2 \setminus Y)$, which 
is independent of $1<\s<2$ by virtue of \reflm(1-2). 
We define $\hat{x}= x/|x|$ for $x\not=0$. 

\bgth \lbth(def-Rext) 
The space $\Rg_{\a,Y}$ is equal to  
\bqn \lbeq(res)
\Big\{\ph(x) 
= \frac{\la\tD\ab, {\bf 1} \ra}{N}  -\frac1{2\pi}  
\sum_{j=1}^N a_j \log |x-y_j| \colon \ab\in \Ker S\tD{S}\cap S\C^N\Big\}.
\eqn  
The function $\ph(x)$ of \refeq(res) satisfies  
\bqn \lbeq(reso-s)
\ph(x)= \frac{\la \tD\ab, {\bf 1}\ra}{N}  + 
\frac1{2\pi}\sum_{j=1}^N \frac{\la \hat{x}, a_j y_j\ra}{|x|} + O(|x|^{-2}) 
\quad (|x|\to \infty).
\eqn 
In particular, $\Rg_{\a, Y} = \{0\}$ if only if $S\tD{S}$ is non-singular 
in $S\C^N$. 
\edth

\bgdf A function $\ph \in \Rg_{\a,Y}$  is called (threshold) resonance 
of $H_{\a,Y}$. 
\ben 
\item[{\rm (1)}] $H_{\a,Y}$ is said be regular at zero if 
$\Rg_{\a,Y}=\{0\}$ otherwise singular at zero .  
\item[{\rm (2)}] $\ph\in \Rg_{\a,Y}$ of \refeq(res) is an   
$s$-wave resonance if $\la \tD\ab, {\bf 1}\ra\not=0$ 
and $p$-wave resonance if $\la \tD\ab, {\bf 1}\ra=0$ but 
$\sum_{j=1}^N a_j y_j\not=0$.
\item[{\rm (3)}] $\ph\in \Rg_{\a,Y}\setminus\{0\}$ 
of in \refeq(reso-s) is an 
eigenfunction of $H_{\a,Y}$ with eigenvalue $0$ 
if $\la \tD\ab, {\bf 1}\ra=0$ and 
$\sum_{j=1}^N a_j y_j=0$.
\een
\eddf 

In the following theorem we use the notation of \refth(Main). 
\bgth \lbth(resonances) 
Suppose $S\tD{S}$ is singular in $S\C^N$. Then: 
\ben
\item[{\rm (1)}] $s$-wave resonances exist if and only if $P\tD T \not=0$. 
\item[{\rm (2)}] All $\ph \in \Rg_{\a,Y}$ are $s$-wave resonances 
if and only if $T{\tD}^2 T$ is non-singular in $T\C^N$. 
\item[{\rm (3)}] Suppose $T{\tD}^2 T$ is singular in ${T\C^N}$. 
Then, $\ph\in \Rg_{\a,Y}$ of \refeq(res) is 
\ben 
\item[{\rm (a)}]   
an $s$-wave resonance if $\ab\in T\C^N \setminus T_p\C^N$. 
\item[{\rm (b)}] a $p$-wave resonance if 
$\ab \in T_p \C^N \setminus T_e \C^N$.
\item[{\rm (c)}] an eigenfunction with eigenvalue $0$ if 
$\ab \in T_e \C^N$. 
\een
\item[{\rm (4)}] The eigenspace of $H_{\a, Y}$ associated with eigenvalue zero 
is the set of all $\ph(x)$ in \refeq(res) with $\ab\in T_e \C^N$. 
\een
\edth 

In virtue of \refth(Main), \refth(def-Rext) and \refth(resonances), 
wave operators are bounded in $L^p(\R^2)$ for all 
$1<p<\infty$ if $H_{\a,Y}$ has no $p$-wave resonances otherwise  
they are bounded only for $1<p \leq 2$.

We briefly record here the result of \cite{CMY} on the threshold behavior 
of $(H_{\a, Y}-z^2)^{-1}$ to show its relation to \refth(Main). 
We refer to \cite{CMY} for more precise result. For Schr\"odinger 
operators with regular potentials, the relation between resonances,  
the threshold behavior of the resolvent and the large time 
behavior of solutions of time dependent Schr\"odinger equation
is extensively studied (see e.g. \cite{JK, Mu, Schlag, JN, EG, EGG}). 

Let $\s>1$ and let $\Bb_\s$ be the  
Banach space of bounded operators from $L^2_{\s}(\R^2)$ to $L^2_{-\s}(\R^2)$. 
Then, the well known limiting absorption principle for $(H_0 -z^2)$ and 
the behavior of the Hankel function imply that  
$(H_{Y,\a}-z^2)^{-1}$ regarded as a 
$\Bb_\s$-valued function of $z \in \C^{+}\setminus\Eg $ can be 
continuously extended to $\Cb^{+}\setminus (\Eg \cup \{0\})$. 
Here and in what follows we use $\lam$ instead of $z$ when we emphasize 
$z$ can also be real not only $z\in \C^{+}$.   

\begin{theorem}[\cite{CMY}] \lbth(zerolimit) 
{\rm (1)} Suppose that $H_{\a,Y}$ is regular at zero, then  
$(H_{Y,\a}-\lam^2)^{-1}$ 
can be extended continuously to $0$. \\[5pt] 
{\rm (2)} Suppose the condition of \refth(Main) {\rm (2)} is satisfied.
Then, $T= \fb \otimes \fb$ for a normalized $\fb\in T\R^N$ and 
\[  
(H_{Y,\a}-\lam^2)^{-1}  = 
a^{-2} g(\lam)\ph \otimes \ph + O(1)  \quad (\lam \to 0), 
\]
where $\ph(x)$ is an $s$-wave resonance defined in \refeq(res) 
with $\fb$ in place of $\ab$.  \\[5pt]
{\rm (3)} 
Suppose the condition of \refth(Main) {\rm (3)} is satisfied.
Then   
\[
(H_{Y,\a}-\lam^2)^{-1}=  -(Ng\lam^{2})^{-1} 
\sum_{j=1}^n  a_j \ph_j(x) \ph_j(y)  + O(\lam^{-2} ) \quad (\lam \to 0), 
\]
where $n= {\rm rank}\, T_p$ and $\ph_j$, $j=1, \dots,, n$ 
are $p$-wave resonances. \\[5pt]
{\rm (4)} Suppose the condition of \refth(Main) {\rm (4)} is satisfied.
Then, 
\begin{align*} 
& (H_{Y,\a}-\lam^2)^{-1}(x,y)\notag \\
& = -  (N \lam^{2})^{-1} \la T_e \hat{N}_{0,Y}(x), 
[T_e\Gg_2(Y) T_e]^{-1} T_e \hat{N}_{0,Y}(y)\ra +  
O(\lam^{-2}g(\lam)^{-1} ). 
\end{align*}
\end{theorem} 

The rest of the paper is devoted to the proof of the lemmas and 
theorems (but not of \refth(zerolimit)). In section 2, we prove results 
on the resonances, \reflm(1-1), \reflm(1-2) and \refth(def-Rext). 
In section 3, we collect results necessary for proving \refth(Main). 
We first recall from \cite{CMY} the stationary 
and the product decomposition formulas for the wave operators 
and the result that the high energy part of the 
$W_{\a,Y}^\pm \chi_{\geq \ep}(|D|)$ is bounded in $L^p(\R^2)$ for all 
$1<p<\infty$. We then examine the result in \cite{CMY} 
on the behavior of $\Ga(\lam)^{-1}$ as $\lam\to 0 $ and  
give an estimate on the Fourier transform of a logarithmic function. 
We prove in section 4 the statement of \refth(Main) 
separately. In virtue of the the high energy results mentioned above 
we prove them for the low energy part $W_{\a,Y}^\pm\chi_{\leq \ep}(|D|)$ 
only. 
Statement (1) is a direct result of the product formula and Mikhlin's theorem 
on Fourier multliplier. Proofs of statements (2) to (5) uses the cancellation 
properties produced by the linear operators $S, T_p$ and $T_e$ of \refth(Main).

\section{Proof of results on resonances } 
In this section we prove \reflm(1-2), \refth(def-Rext) and \refth(resonances).  
The Fourier transform $\hat u(\xi) = {\Fg}u(\xi)$ is 
defined by  
\[
\Fg u(\xi)= \frac1{2\pi}\int_{\R^2} e^{-ix\xi}u(x) dx. 
\]

\paragraph{Proof of \reflmb(1-2).}  
Define for $\ab \in \C^N$ and $z\in \C^{+}$,  
\bqn 
h_z(x) \stackrel{\mathrm {def}}
{=} \la \ab, \hat{N}_{0,Y}(x)-\hat{\Gg}_{z,Y}(x) \ra .
\lbeq(h1) 
\eqn 
We have $h_z(x) \in H^2_{-\s}(\R^2)$ for $1<\s<2$ 
because \refeq(hankel-1) implies the $\log$-singularities at $x=y_j$ of 
cancel and it grows only logarithmically as $|x|\to \infty$. 
Moreover, ${N}_0(x-y_j)- {\Gg}_{z}(x-y_j)= g(z)+ O(g(z)^2|x-y_j|^2)$ 
as $x\to y_j$ and, the definitions of $\Ga_{\a,Y}(z)$ and $\tD$ imply   
\begin{align}
h_z(y_j)& = \sum_{k=1 }a_k 
\left(\big(-\frac1{2\pi}\log|y_k-y_j|- {\Gg}_z(y_j-y_k)\big)\hat{\d}_{jk}
- g(x)\d_{jk} \right) \lbeq(h2)  \\
& = [(\Ga_{\a,Y}(z)-\tD)\ab ]_j, \quad j=1, \dots, N  \lbeq(h3).
\end{align}
(a) We first show that $\ph(x)$ defined by \refeq(phlog) with 
$\ab \in \Ker\, S\tD$ is a solution of 
$H_{\a,Y}^{-\s} \ph =0$. Let $C = N^{-1}\la \tD\ab, {\bf 1}\ra$. 
By virtue of \refeq(h1) 
\bqn \lbeq(ph-ker)
\ph(x) = v_z(x) + \la \ab, \hat{\Gg}_{z,Y}(x) \ra \ \mbox{with}\    
v_z(x)= h_z(x) + C \in H^2_{-\s}(\R^2)
\eqn 
and, we have $\Ga_{\a,Y}(z)\ab=\hat{v}_{z,Y}$ because \refeq(h3) and 
$S\tD\ab=0$ or $\tD\ab\in P\C^N$ imply 
$\hat{v}_{z,Y}- \Ga_{\a,Y}(z)\ab= -\tD\ab + C{\bf 1}
=(-{N^{-1}}\la \tD\ab, {\bf 1}\ra -C){\bf 1} =0$. 
It follows $\ph \in D(H_{\a,Y}^{-\s})$. Moreover, \refeq(def-of-hweight) 
and \refeq(ph-ker) imply $H^{-\s}_{\a,Y}\ph(x)=0$ because 
\[
(H^{-\s}_{\a,Y}-z^2)\ph(x) = (-\lap -z^2)v_z(x) 
= -z^2 (C +\la \ab, \hat{N}_{0,Y}(x)\ra) = -z^2 \ph(x).
\]
(b) Assume conversely that $\ph(x)$ satisfies $H_{\a,Y}^{-\s}\ph=0$. 
We show that $\ph(x)$ is necessarily of the form \refeq(phlog) with 
$\ab$ such that $S\tD\ab=0$. Since $\ph \in D(H^{-\s}_{\a,Y})$,
for $z\in C^{+}\setminus \Eg$ there must exists $v_z \in H^2_{-\s}(\R^2)$ 
such that  
\bqn \lbeq(phx) 
\ph(x) = v_z (x) + \la \ab_z, \hat{\Gg}_{z,Y}(x)\ra \ \mbox{where} \  
\ab_z= \Ga_{\a, Y}(z)^{-1}\hat{v}_{z,Y} 
\eqn 
and $(H_{\a,Y}^{-\s}-z^2) \ph = (-\lap -z^2)v_z$ or  
$-z^2\ph(x) = (-\lap -z^2)v_z(x)$. hence, 
\bqn \lbeq(2-6)
-\lap v_z(x) = z^2(v_z(x) -\ph(x))= -z^2 \la \ab, \hat{\Gg}_{z,Y}(x)\ra.
\eqn 
We observe that 
$(\ab_z)_j = -\lim_{x\to y_j} (2\pi)(\log |x-y_j|)^{-1}\ph(x)$ 
because $v_z\in H^{2}_{-\s}(\R^2)$ is continuous and $\ab_z$ is 
independent of $z$. Thus, we write $\ab$ 
for $\ab_z$ and define $h_z(x)$ by \refeq(h1) with this $\ab$. We will
show $v_z(x) =  h_z(x) + C$ for a constant $C$ and, 
hence $\ph(x)= \la \ab, \hat{N}_{0,Y}(z) \ra + C$, or it must be of the 
form in \refeq(phlog). Indeed, it follows from \refeq(2-6) that 
\[
-\lap \left(v_z+ \la \ab, \hat{\Gg}_{z,Y}\ra\right)
= \la \ab, (-\lap -z^2)\hat{\Gg}_{z,Y}\ra 
= \la \ab, \d_{Y}\ra
= \la \ab, -\lap \hat{N}_{0,Y}\ra
\]
and $0= -\lap \left(v_z(x)+ 
\la \ab,(\hat{\Gg}_{z,Y}(x)-\hat{N}_{0,Y}(x)\ra \right)
=-\lap(v_z(x) -h_z (x))$. 
Thus, $v_z(x) - h_z(x)$ must be a harmonic polynomial which belogns to 
$H^2_{-\s}(\R^2)$ and, hence $v_z(x)-h_z(x) = C$. For determining $C$,
we recall \refeq(h3), which implies 
\[
C{\bf 1}= 
\hat{v}_{z,Y}- \hat{h}_{z,Y} = \Ga_{\a, Y}(z)\ab - (\Ga_{\a, Y}(z)-\tD)\ab 
= \tD\ab.  
\]
It follows that $\ab$ must be such that 
$\tD\ab\in P\C^N$ and $C=N^{-1}\la \tD\ab, {\bf 1}\ra$. \qed

\paragraph{Proof of \refthb(def-Rext)} 
The function $\ph(x)= C + \la \ab, \hat{N}_{0,Y}(x)\ra$, $S\tD\ab={\bf 0}$ 
of \refeq(phlog) is evidently continuous in $\R^2 \setminus Y$. 
It is also evident that $\ph(x)$ is bounded near infinity if 
and only if $\la {\bf 1}, \ab\ra=0$ or $\ab\in S\C^N$. This implies 
that $\Rg_{\a,Y}$ is given by \refeq(res). 
The relation \refeq(reso-s) is evident if 
$\la {\bf 1}, \ab\ra=a_1+ \cdots+ a_N=0$. \qed

\paragraph{Proof of \refthb(resonances)} 
(1) If $H_{\a,Y}$ has an $s$-wave resonance, then 
there must exists non-zero $\ab \in T\C^N= S\C^N \cap \Ker\, S\tD$ 
such that $P\tD\ab\not=0$, hence $P\tD{T}\not=0$. Conversely, if  
$P\tD{T}\not=0$, then there exists $\ab \in T\C^N$ such that 
$P\tD\ab\not=0$ and $\ph(x)$ 
defined by \refeq(phlog) with this $\ab$ is an $s$-wave resonance. 

(2) If $T{\tD}^2 T$ is non-singular, 
$\Ker T{\tD}^2 T = \Ker \tD T =\{0\}$ since $\tD$ is real symmetric 
and, for 
$\ab\in T\C^N \setminus\{0\}$ we have $P\tD \ab = \tD\ab \not=0$. 
Thus all resonances are $s$-wave resonances. 
If $T{\tD}^2 T$ is singular on the other hand then, $\tD \ab =0$ 
for an $\ab \in T\C^N$ which trivially satisfies $P\Dg T\ab =0$ 
and, \refeq(phlog) with this $\ab$ produces a $p$-wave resonance. 

(3a) If $\ab \in T\C^N$ is such that $T{\tD}^2 T\ab\not=0$, 
then $\tD \ab \not=0$ and $P\tD \ab \not=0$ as $S\tD \ab=0$. Thus,  
$\ph(x)$ produced by $\ab$ by \refeq(phlog) is an $s$-wave resonance. 

(3b,c) For $\ab \in T_p \C^N\setminus \{0\}$, we have ${\tD}\ab={\tD}T\ab=0$ 
and $P\Dg\ab=0$ trivially. If $T_p \hat{\Gg}_1(Y) T_p \ab\not=0$, 
$\la \hat{\Gg}_1(Y) \ab, \ab \ra = \la Y, \ab \ra^2 \not=0$  
and $\ph(x)$ produced by $\ab$ by \refeq(phlog) is a $p$-wave resonance; 
but if $T_p \hat{\Gg}_1(Y) T_p \ab=0$, $\ph(x) = O(|x|^{-2})$ as $|x|\to \infty$ 
and it is an eigenfunction.  

(4)If $\ab \in T_e \C^N\subset T_p \C^N$, then we have seen in (3b) that 
$\la \tD\ab, {\bf 1}\ra=0$ 
and $\la Y, \ab \ra={\bf 0}$. Hence $\ph(x)$ is an eigenfunction. 
Conversely, eigenfunction is also of the form \refeq(phlog) with $\ab \in T\C^N$ 
by virtue of \refth(def-Rext) and $\ab$ must further satisfy 
$\la \tD\ab, {\bf 1}\ra=0$ and $\la Y, \ab \ra={\bf 0}$. 
The former implies $\tD \ab=0$, hence $T\tD^2 T\ab=0$ and 
$\ab \in T_p\C^N$ and the latter $T_p \hat{\Gg}_1(Y)T_p \ab =0$, 
which implies $\la \ab, T_p \hat{\Gg}_1(Y)T_p \ab\ra =0$ 
and $T_p \hat{\Gg}_1(Y)T_p\ab=0$ since $T_p\hat{\Gg}_1(Y)T_p$ 
is non-negative on $T_p\C^N$. Thus, we have $\ab \in T_e \C^N$.  
\qed

\section{Preliminary for the proof of \refthb(Main)} 
In this section, we collect several lemmas which we use for proving  
\refth(Main), some of which are well known and are recorded 
for readers convenince. The strength $\a$ and points of interaction $Y$ 
will be fixed hereafter and will be often omitted from various formulas, 
e.g. $\hat\Gg_\lam(x)= \hat{\Gg}_{\lam,Y}$.  
We prove \refth(Main) for $W^{-}$. We have $W^{+}= \Cg W^{-}\Cg^{-1}$ 
by the complex conjugation $\Cg u(x) = \overline{u(x)}$ and the result 
for $W^{+}$ immediately follows from that for $W^{-}$. 
We set  
\bqn \lbeq(Dgast-def)
\Dg_\ast = 
\{u \in \Sg(\R^2) \colon \hat{u}\in C_0^\infty(\R^2 \setminus\{0\}) \}.
\eqn 
The space $\Dg_\ast$ is dense in $L^p(\R^2)$ for any $1<p<\infty$. 
For Borel functions 
$f$, $f(D)$ is the Fourier multiplier by $f(\xi)$: 
$f(D)u(x)= \Fg^{-1}(f(\xi)\hat{u}(\xi))(x)$. 
\[
\tGa(\lam) {=} {\Ga(\lam)}^{-1}
\]
and $\tGa(|D|)=(\tGa_{jk}(|D|))$ is the operator matrix of 
Fourier multipliers  
$\tGa_{jk}(|D|)$. For $y \in \R^2$, $\t_y u(x)=u(x-y)$ is translation by $y$. 
We define the Riesz transform (\cite{Stein}) 
$R=\begin{pmatrix} R_1 \\ R_2 \end{pmatrix}$ by  
$R_l= R_l(D)= \Fg^{-1}(\xi_l/|\xi|)\Fg$, $l=1,2$ and 
for a vector $a\in \R^2$, $\la a, R \ra u= a_1 R_1 u + a_2 R_2 u$. 
$\la a, R \ra$ is a bounded operator in $L^p(\R^2)$ for $1<p<\infty$. 

\subsection{Product decomposition of wave operators}

\paragraph{Stationary representation. }
The following representation of the wave operator $W^{+}$  
via the resolvent kernel $\Gg_{\lam}(x)$ may be proved by following 
the argument of the proof of Lemma 3.2 of \cite{DMSY} 
for the three dimensional case. 

\bglm \lblm(stationary)  Wave operator $W^{-}$ may be repesented in the form 
\bqn 
W^{-} u= u+ 
\frac1{\pi{i}}\int_0^\infty\int_{\R^2} \lam 
\la \tGa(\lam)\hat{\Gg}_{\lam}(x),  
\hat{\Gg}_{\lam}(y)-\hat{\Gg}_{-\lam}(y)\ra{u(y)}dy d\lam   \lbeq(Stat-2a)
\eqn 
for $u\in \Dg_{\ast}$. Equivalently 
$W^{-} u= u + \sum_{j,k=1}^N \tau_{y_j} \W_{jk}\tau_{y_k}^\ast $ where 
\bqn \lbeq(second)
\W_{jk}u(x) = \frac1{\pi{i}}\int_0^\infty \lam \tGa_{jk}(\lam)\Gg_{\lam}(x) 
\left(\int_{\R^2}(\Gg_{\lam}(y)-\Gg_{-\lam}(y))u(y)dy \right) d\lam .
\eqn
\edlm 

\paragraph{Decomposition formula} 
The following is a slight modification of the 
result of Lemma 4.3 and 4.4 of \cite{CMY}. 
Define the operator $K$ by 
\bqn \lbeq(DEFK)
Ku(x) = \frac1{\pi^2}\int_0^\infty \Gg_{\lam}(x)\lam \left( \int_{{\mathbb S}^1}
\Fg{u}(\lam\w)d\w \right)d\lam.
\eqn 

\bglm \lblm(prod) {\rm (1)} For $j,k=1, \dots, N$, 
$\W_{jk}$ is the product of $\tGa_{jk}(|D|)$ and $K$: 
\bqn \lbeq(prod)
(\W_{jk}u)(x)= (K \circ \tGa_{jk}(|D|))u(x), \quad u \in \Dg_{\ast}.
\eqn 
{\rm (2)} $K$ is a singular integral operator: 
\bqn \lbeq(Kxy)
Ku(x) = \lim_{\ep \downarrow 0}\frac{2}{\pi^2 i}
\int_{\R^2} \frac{u(y)dy}{x^2-y^2 + i\ep}.
\eqn 
{\rm (3)}  For any $1<p<\infty$, there exists a constant $C_p>0$ such that 
\bqn \lbeq(Kest)
\|Ku\|_{L^p(\R^2)}\leq C_p \|u\|_{L^p(\R^2)}, \quad u \in \Dg_{\ast}.
\eqn
\edlm

\paragraph{Mikhlin multiplier} 
We recall the well-known result on the Fourier multiplier 
which will be very often used in what follows. 

\begin{lemma}[Mikhlin] Let $m \in C^2(\R^2 \setminus\{0\})$ 
satisfy $|\pa_{\xi}^\a m(\xi)|\leq C |\xi|^{-|\a|}$ for $|\a|\leq 2$. 
Then, $m(D) \in \Bb(L^p(\R^2))$ for $1<p<\infty$. \lblm(Mikhlin)
\end{lemma} 
We often say that $m(\lam)$ is a {\it good multiplier}  when it 
satisfies the condition of \reflm(Mikhlin).

\subsection{High energy part $W^{-}\chi_{\geq \ep}(H_0)$.} 

The $L^p$ property of the high energy part of $W^{-}$ does not 
depend on the small $z$ behavior of $(H_{\a,Y}-z^2)^{-1}$ 
and the following is proved in \cite{CMY}. In what follows 
$\chi$ will stand for the real function $\chi\in C_0^\infty(\R)$ 
which satisfies  
\bqn \lbeq(chi-def)
\mbox{$\chi(\lam)=1$ for $|\lam|<1$ and 
$\chi(\lam)=0$ for $|\lam|>2$}
\eqn 
and, for $\ep>0$, we define 
\bqn \lbeq(chiep-def)
\chi_{\leq \ep}(\lam)= \chi(\lam/\ep), \quad 
\chi_{\geq \ep}(\lam)=1-\chi_\ep(\lam).
\eqn  
When $\ep>0$ is fixed, then we often write 
$\chi_{\leq}(\lam)= {\chi}_{\leq \ep}(\lam)$ and 
$\chi{\geq}(\lam)={\chi}_{\geq \ep}(\lam)$ omitting $\ep>0$.

\begin{theorem}  \lbth(high) For any $\ep>0$,  
$W_{\a,Y}^{-} (1-\chi_{\leq \ep}(|D|))$ is bounded 
from $L^p(\R^2)$ to itself for any $1<p<\infty$.
\end{theorem}

We say for simplicity that an operator is a {\it good operator} 
if it is bounded from $L^p(\R^2)$ to itself for any $1<p<\infty$. 
By virtue of \refth(high), we need consider only the low energy part 
$W_{\rm low}= W^{-}\chi_{\leq \ep}(|D|)$ in what follows. 

\subsection{Expansion of $\Ga(\lam)^{-1}$} 

For proving \refth(Main) we need precise information on 
the behavior of $\tGa(\lam)= \Ga(\lam)^{-1}$ as $\lam \downarrow 0$. 
We have already obtained some results in \cite{CMY} and \cite{CMYE}, 
however, because we shall need some more precise results and those 
which are buried in proofs, we have decided to completely redo it. 
The low energy behavior of $\Ga(\lam)^{-1}$ is different depending on the 
conditions of statements of \refth(Main) 
and we split the subsection into five paragraphs accordingly. 
We remark that the conditions in each steps are mutually exclusive.

In what follows,  for two functions $f(\lam)$ and $h(\lam)$ on $(0,\infty)$,  
$f(\lam)= O(h(\lam))$ means that for $j=0,1, \dots$ that
$f^{(j)}(\lam)\absleq C_j h^{(j)}(\lam)$ is satisfied for 
$0<|\lam|<\ep_j$ for some $\ep_j>0$ and $C_j>0$.  
Hereafter we shall indiscrimately denote by 
$M(\lam)$ a good multiplier which may differ at each appearance  

We shall repeatedly use the following lemma due to Jensen and Nenciu \cite{JN}. 
The following trivial identities for matrices will be frequently used:
\begin{gather} 
(1+X)^{-1}= 1-X(1+X)^{-1}= 1- (1+X)^{-1}X, \lbeq(X-1)\\  
(1+X)^{-1}= 1- X + X(1+X)^{-1}X. \lbeq(X-2)
\end{gather} 

\begin{lemma} \lblm(JN) 
Let $A$ be a closed operator in a Hilbert space $\Hg$ 
and $S$ a projection. Suppose $A+S$ has a bounded inverse. 
Then, $A$ has a bounded inverse if and only if 
\[
B= S - S(A+S)^{-1}S 
\]
has a bounded inverse in $S\Hg$ and, in this case, 
\bqn \lbeq(JN-1)
A^{-1}= (A+S)^{-1}+ (A+S)^{-1}SB^{-1}S (A+S)^{-1}.
\eqn 
\end{lemma}

We also use the well known Feshbach formula but only at the last step. 
In what follows we use the notation of \refth(Main) 
and omit the variable $\lam$ from various functions 
when no confusion is feared. Identity matrices of various subspaces 
are indiscrimately denoted by $1$ and orthogonal 
projections $P$ in subspaces $V$ will be often regarded as the projection 
$P \oplus 0$ in the full space $\C^N= V \oplus V^\perp$.

\paragraph{Step 1.}  Define $A(\lam)$ by 
\[
\Ga(\lam) = -NgA(\lam)
\]
and 
\bqn 
F {=} - N^{-1}\tD, \quad 
\Rg  = \Gg_1(Y) + g(\lam)^{-1}\Gg_2(Y) . \lbeq(A-a) 
\eqn 
Then, \refeq(expansion-1) implies  
\bqn 
A(\lam)+ S  = 1 + g^{-1} F + \Mg_0(\lam), \quad 
\Mg_0(\lam){=} \lam^2 \Rg + O(\lam^4). \lbeq(A) \\
\eqn 
Thus, $A+S$ is non-singular for small $\lam>0$ and by virtue of \refeq(X-2)   
\begin{align} \lbeq(A+Sinv)
& (A+ S)^{-1}= (1+ g^{-1}F)^{-1}(1+ \Mg_0(1+ g^{-1}F)^{-1})^{-1} \\
& = (1+ g^{-1}F)^{-1} - (1+ g^{-1}F)^{-1}\Mg_0(1+O(\lam^2))(1+ g^{-1}F)^{-1}.  
\lbeq(A+Sinv-a)
\end{align}
In particular $(A+S)^{-1}$ is a good multiplier and   
\bqn \lbeq(A+S-Sinter)
(A+ S)^{-1}S  = S + O(g^{-1}), \quad 
S (A+ S)^{-1}  = S + O(g^{-1}).
\eqn  
In view of \reflm(JN), we define $B(\lam)$ and $A_1(\lam)$ by 
\bqn 
B {=} S- S(A+S)^{-1}S = - N^{-1}g(\lam)^{-1} A_1. \lbeq(B-0) 
\eqn 
We substitute \refeq(A+Sinv-a) for $(A+S)^{-1}$ in \refeq(B-0). 
Since 
\bqn \lbeq(L-def)
L \stackrel{\rm def}{=}-Ng(S-S(1+ g^{-1}F)^{-1}S) 
= S{\tD}S + g^{-1}N^{-1} S\tD^2(1+ g^{-1}F)^{-1}S
\eqn 
we have 
\begin{align}
A_1 & =L - g SN(1+ g^{-1}F)^{-1}\Mg_0(1+O(\lam^2))(1+ g^{-1}F)^{-1}S
\lbeq(A1-exp) \\
& {=} S\tD{S}+ S\tD^2(gN)^{-1}(1+g^{-1}F)^{-1}S - g\lam^2 \Rg_1  
\lbeq(G-3)
\end{align}
where we defined $\Rg_1=O(1)$ by  
\bqn 
\Rg_1 \stackrel{\rm def}{=}  N S (1+g^{-1}F)^{-1}
(\Rg+ O(\lam^2))(1+g^{-1}F)^{-1}S . \lbeq(tR-def)
\eqn 

We have the following result.   
\bglm \lblm(Step1)
If $S{\tD}S$ is non-singular in $S\C^N$, then  
$\Ga(\lam)^{-1}$ is a good multiplier. 
\edlm 
\bgpf \refeq(G-3) implies $A_1= S\tD{S}+ O(g^{-1})$. Hence,  
$A_1^{-1}$ exists in $S\C^N$ and it is a good multiplier. 
It follows by virtue of \refeq(JN-1) of \reflm(JN) that 
\bqn 
\Ga(\lam)^{-1} = -N^{-1}g^{-1}(A+S)^{-1} + 
(A+S)^{-1}S A_1^{-1}S (A+S)^{-1}. \lbeq(G-1) 
\eqn  
Since $(A+S)^{-1}$ is a good multilier, \refeq(G-1) 
implies that $\Ga(\lam)^{-1}$ is also a good multiplier. 
\edpf 

\paragraph{Step 2.} 
Suppose next $S{\tD}S$ is singular in $S\C^N$ and let $T$ be 
the orthogonal projection to $\Ker_{S\C^N} S{\tD}S$.  
>From \refeq(G-3) we have 
\[ 
L+T =(S\tD{S}+T)+ N^{-1}g^{-1}S\tD(1+g^{-1}F)^{-1}\tD{S}
\]
and $S\tD{S}+T$ is clearly invertible in $S\C^N$. Hence  
\bqn 
L+T= (1+  g^{-1}\tilde{F})(S\tD{S}+T).
\eqn 
where $\tilde{F}= O(1)$ is defined by   
\bqn 
\tilde{F} \stackrel{\rm def}{=} 
N^{-1}S\tD(1+g^{-1}F)^{-1}\tD{S}(S\tD{S}+T)^{-1}, 
\lbeq(tFdef)
\eqn 
It follows that $L+T$ is also invertible in $S\C^N$ for small $\lam>0$ 
and        
\begin{multline} 
(L+T)^{-1} = (S\tD{S}+T)^{-1} - g^{-1}(S\tD{S}+T)^{-1}\tilde{F} \\
+ g^{-2}(S\tD{S}+T)^{-1}\tilde{F}^2 (1+g^{-1}\tilde{F})^{-1} 
\lbeq(L+T)
\end{multline}   
by virtue of \refeq(X-2). Then, \refeq(A1-exp) implies that 
$A_1+ T$ is also invertible in 
$S\C^N$ and 
\begin{align} \lbeq(A1+T)
& (A_1+ T)^{-1}= (L+T)^{-1}
(1- g\lam^2 \Rg_1(L+T)^{-1})^{-1} \\
& = (L+T)^{-1}+ g\lam^2(L+T)^{-1}
(\Rg_1+ O(\lam^2g))(L+T)^{-1} . 
\lbeq(3-64)
\end{align} 
\refeq(L+T) and \refeq(3-64) imply that 
$(A_1+ T)^{-1}$ in $S\C^N$ is a good multiplier and that 
\bqn 
(A_1(\lam)+ T)^{-1} T = T + O(g^{-1}), \quad 
T(A_1(\lam)+ T)^{-1} = T + O(g^{-1}).  \lbeq(A1-T-inter)
\eqn 
For studying $A_1^{-1}$ by using \reflm(JN), define $B_1$ in $T\C^N$ by 
\bqn \lbeq(B1-def)
B_1 {=} T - T(A_1+ T)^{-1}T.
\eqn 
Inserting \refeq(3-64) into \refeq(B1-def), we have 
\bqn 
B_1= T- T(L+T)^{-1}T - g\lam^2 \Rg_2,  \lbeq(B1) 
\eqn 
where we defined $\Rg_2$ by 
\bqn 
\lbeq(R2-def)
\Rg_2 \stackrel{\rm def}{=} T(L+T)^{-1}(\Rg_1 + O(g\lam^2))(L+T)^{-1}T.  
\eqn  
Since $T(S\tD{S}+T)^{-1}= T$, \refeq(L+T) and \refeq(tFdef) imply 
that 
\begin{multline*}  
T- T(L+T)^{-1}T \\
= 
g^{-1}T\tilde{F}T - g^{-2}T \tilde{F}^2({{1}}
+g^{-1}\tilde{F})^{-1}T 
{=}N^{-1}g^{-1} (T\tD^2{T}+ g^{-1}F_2) ,
\end{multline*}
where $F_2$ is defined by 
\bqn \lbeq(F2-def)
F_2(\lam)= N^{-1}T\tD^2({{1}}+g^{-1}F)^{-1}\tD{T}- N
\tilde{F}^2({{1}}+g^{-1}\tilde{F})^{-1}T
\eqn 
should be obvious. 
With these definition $B_1$ becomes 
\bqn \lbeq(A2-def)
B_1 = N^{-1}g^{-1} A_2, \ \ A_2 = 
T\tD^2 T + g^{-1}F_2 - g^2 N \lam^2  \Rg_2 .
\eqn 

\bglm \lblm(Ga-pwave)
Suppose $T{\tD}^2 T$ is non-singular in $T\C^N$. Then $\dim T\C^N=1$, 
${\rm rank}\, T\tD^2 T=1$ and $H_{\a,Y}$ has an $s$-wave resonance only. 
Moreover, 
\bqn  
\Ga(\lam)^{-1} =  N g T (T\tD^2 T)^{-1} T +  M(\lam)  .    \lbeq(S2-extra)
\eqn 
where we wrote $T$ for $ST$ and $TS$ in \refeq(S2-extra).
\edlm 
\bgpf The first part of the lemma is proved in \cite{CMYE}. 
If $T\tD^2 T$ is non-singular in $T\C^N$, then \refeq(A2-def) implies 
$A_2$ and, hence $B_1$ are non-singular in $T\C^N$,  
\bqn \lbeq(B1-inv)
A_2^{-1} = (T\tD^2 T)^{-1} + O(g^{-1}), \quad 
B_1^{-1} = Ng((T\tD^2 T)^{-1} + O(g^{-1}))
\eqn
and by virtue of \reflm(JN) 
\bqn \lbeq(A1-inv)
A_1^{-1} = (A_1 + T)^{-1}+ (A_1 + T)^{-1} T B_1^{-1}T (A_1 + T)^{-1}.
\eqn 
Combining \refeq(G-1) and \refeq(A1-inv), we have  
\begin{align} 
& \Ga(\lam)^{-1} = -N^{-1} g^{-1} (A+S)^{-1} + 
(A+S)^{-1}S (A_1+ T)^{-1} S (A+S)^{-1} \lbeq(Step1-1)  \\
& \hspace{1cm}+ Ng (A+S)^{-1}S (A_1+ T)^{-1}T A_2^{-1} 
T(A_1+T)^{-1} S (A+S)^{-1}. \lbeq(Step1-1a) 
\end{align}
As $(A+S)^{-1}$ and $(A_1+ T)^{-1}$ are good multiplies as has been proved 
previously, the two terms in \refeq(Step1-1) are good multipliers. 
\refeq(B1-inv), \refeq(A1-T-inter) and \refeq(A+S-Sinter) implies 
that \refeq(Step1-1a) is equal to the sum of 
$N g ST (T\tD^2 T)^{-1} TS$ and a good multiplier. This proves \refeq(S2-extra). 
\edpf

\paragraph{Step 3.}  
We next assume that $S\tD{S}\vert_{S\C^N}$ and $T\tD^2 T\vert_{T\C^N}$  are 
both singular. Let $T_p$ be the orthogonal projection onto 
$\Ker\, T\tD^2 T$ in $T\C^N$.  
Recall that we irrespectively write $M(\lam)$ 
for a good multiplier. Recalling \refeq(A2-def), we define 
\bqn \lbeq(tildeA2-def)
\tilde A_2 \stackrel{\rm def}{=}  T\tD^2 T +  g^{-1}F_2 \ \
\mbox{\rm so\ that } \ A_2 = \tilde A_2 - g^2 \lam^2 N {\Rg}_2\, .
\eqn 
It is evident that $T\tD^2 T + T_p$ is non-singular in $T\C^N$. 
Hence both of $\tilde A_2+T_p$ and $A_2+ T_p$ are invertible in $T\C^N$ 
for small $\lam>0$ and   
\begin{align}
(A_2+ T_p)^{-1} & = (\tilde A_2 + T_p)^{-1}
+ g^2 \lam^2 N (\tilde A_2 + T_p)^{-1}\Rg_2(1+ O(\lam^2 g^2))
(\tilde A_2 + T_p)^{-1}\notag  \\
& = (T\tD^2 T + T_p)^{-1} + O(g^{-1}) 
\lbeq(A2Tp)
\end{align}
In particular $(A_2+ T_p)^{-1}$ is a good multiplier.

\bglm \lblm(4-3) 
{\rm (1)} The projection $T_p$ annihilates all of $\tD, \tilde{F}, L, F_2 $ and 
$\tilde{A}_2$:  
\bqn  \lbeq(zero)
\br{lll} 
\tD{T}_p= T_p \tD =0, \quad & T_p \tilde{F}=\tilde{F}T_p=0 ,& T_p L = LT_p=0, 
\\
T_p F_2= F_2 T_p=0, \quad & \tilde A_2 T_p = T_p \tilde A_2 =0. &  
\er 
\eqn
{\rm (2)} We have the following identities:
\begin{gather*} 
(A+S)^{-1}S (A_1+ T)^{-1}T (A_2+ T_p)^{-1}T_p 
= T_p + \lam^2 g^2 S M T_p + \lam^2 g M T_p. \\
 T_p (A_2+ T_p)^{-1} T (A_1+ T)^{-1}S (A+S)^{-1} 
= T_p + \lam^2 g^2 T_p M S + \lam^2 g T_p M. 
\end{gather*}
\edlm 
\bgpf (1) Since $\tD$ is real symmetric, we have 
$\Ker T\tD^2 T= \Ker \tD{T}$ and $\tD{T}_p= T_p \tD =0$. 
Other identities of (1) follows from this immediately. 
It follows from \refeq(tildeA2-def) that 
$T_p(A_2+T_p)= T_p+ T_p O(\lam^2g^2)$,  
$(A_2+T_p)T_p= T_p+ O(\lam^2g^2)T_p$ and hence 
\bqn 
T_p (A_2+T_p)^{-1}= T_p + T_p O(\lam^2g^2), \ \  
(A_2+T_p)^{-1}T_p = T_p +O(\lam^2 g^2)T_p.   \lbeq(A2-Tpinter)
\eqn 
Likewise we have from \refeq(G-3) and \refeq(A) that 
\begin{gather}
(A_1+ T)^{-1} T_p = T_p + O(\lam^2 g)T_p, \ \  
T_p (A_1+ T)^{-1}= T_p + T_p O(\lam^2 g).   \lbeq(A1-Tpinter) \\ 
(A+S)^{-1}T_p = T_p + O(\lam^2)T_p, \ \  
T_p (A+S)^{-1} = T_p + T_p O(\lam^2). \lbeq(A+S-Tpinter)
\end{gather} 
Then, we apply \refeqss(A2-Tpinter,A1-Tpinter,A+S-Tpinter) consecutively 
in this order and then \refeq(A1-T-inter) and \refeq(A+S-Sinter) to obtain  
\begin{align*}
&(A+S)^{-1}S (A_1+ T)^{-1}T (A_2+ T_p)^{-1}T_p  \\
& = (A+S)^{-1}S (A_1+ T)^{-1}T_p  
+ (A+S)^{-1}S (A_1+ T)^{-1}T O(\lam^2 g^2)T_p \\
& =(A+S)^{-1}S (T_p + O(\lam^2 g)T_p) +  
(A+S)^{-1}S (T + O(g^{-1})T) O(\lam^2 g^2)T_p  \\
& = T_p + O(\lam^{2})T_p+ SO(\lam^2 g)T_p+ SO(\lam^2 g^2)T_p+ O(\lam^2 g)T_p.
\end{align*} 
This yields the first of (2). The second is the conjugate of the first.
\edpf 

We study $A_2^{-1}$ via \reflm(JN) and we define   
\bqn 
B_2 = T_p - T_p (A_2+ T_p)^{-1}T_p . \lbeq(B2-1)
\eqn
By virtue of \refeq(zero) we have $T_p (\tilde A_2 + T_p)^{-1} 
= (\tilde A_2 + T_p)^{-1}T_p = T_p$ and 
\bqn
T_p(A_2+T_p)^{-1}T_p = T_p + N g^2 \lam^2  T_p 
({\Rg}_2+O(\lam^2 g^2)) T_p. \lbeq(S5-3)
\eqn
Equations \refeq(zero) impy 
$T_p (L+T)^{-1}= (L+T)^{-1}T_p = T_p $  
and, recalling the definitions \refeq(R2-def) and \refeq(tR-def), we obtain    
\bqn \lbeq(B-ob)
T_p \Rg_2 T_p = NT_p (\Rg + O(g\lam^2)) T_p= 
NT_p(\Gg_1(Y) + g^{-1}\Gg_2(Y)+O(g\lam^2))T_p.
\eqn 
It follows by applying \refeq(S5-3) and \refeq(B-ob) to \refeq(B2-1) that 
\bqn \lbeq(A3def)
B_2
= -N^2 g^2 \lam^2 A_3, \ \ 
A_3=T_p {\Gg}_1(Y) T_p + g^{-1}T_p {\Gg}_2(Y) T_p + T_p O(\lam^2 g^2)T_p 
\eqn 

\bglm \lblm(p-wave)
Suppose $T{\tD}^2 T$ is singular in $T\C^N$ and let $T_p$ be 
the projection onto $\Ker\, T{\tD}^2 T$ in $T\C^N$. Suppose 
$T_p \hat{\Gg}_1(Y) T_p$ is non-singular in $T_p \C^N$. Then, 
denoting $STT_p$ and $T_p T S$ simply by $T_p$, 
\bqn \lbeq(Gam-S3)
\Ga(\lam)^{-1} = 
-N^{-1}g^{-1}\lam^{-2}T_p
(T_p {\Gg}_1(Y) T_p + g^{-1}T_p {\Gg}_2(Y) T_p)^{-1}T_p 
+ g SM S + M.  
\eqn 
\edlm 
\bgpf If $T_p{\Gg}_1(Y) T_p$ is non-singular in $T_p \C^N$, then $B_2$ and 
$A_3$ are invertible in $T_p \C^N$ by virtue of \refeq(A3def) and  
\reflm(JN) implies 
\bqn 
A_2^{-1}= (A_2+T_p)^{-1} -N^{-2}g^{-2} \lam^{-2}
(A_2+T_p)^{-1}T_p A_3^{-1} T_p (A_2+T_p)^{-1}.
\lbeq(A2-1) 
\eqn 
We substitute  \refeq(A2-1) for $A_2^{-1}$ of \refeq(Step1-1a). 
Then, modulo $M(\lam)$ which is produced by \refeq(Step1-1), 
$\Ga(\lam)^{-1}$ 
is equal to   
\begin{align}
& Ng (A+S)^{-1}S (A_1+ T)^{-1}T (A_2+T_p)^{-1} T(A_1 +T)^{-1} S (A+S)^{-1} 
\lbeq(Step3-2)  \\
& - N^{-1}g^{-1}\lam^{-2} (A+S)^{-1}S (A_1+ T)^{-1}T(A_2+T_p)^{-1}  \notag \\
& \hspace{3cm} \times 
T_p A_3^{-1} T_p (A_2+T_p)^{-1} T(A_1 +T)^{-1} S (A+S)^{-1}. \lbeq(Step3-3) 
\end{align}
Substituting \refeq(A2Tp) that 
$(A_2+ T_p)^{-1}=  (T\tD^2 T + T_p)^{-1} + O(g^{-1})$ 
for \refeq(Step3-2) and applying the relations  
\refeq(A+S-Sinter) and \refeq(A1-T-inter), we obtain    
\bqn \lbeq(L2)
\refeq(Step3-2)=Ng ST (T\tD^2 T + T_p)^{-1} T S +M(\lam).
\eqn 
For studying \refeq(Step3-3) we introduce a short hand notation 
\[
G_1 \stackrel{{\rm def}}{=} T_p {\Gg}_1(Y) T_p, \quad 
G_2 \stackrel{{\rm def}}{=} T_p {\Gg}_2(Y) T_p, \quad 
\tilde {\Rg}\stackrel{{\rm def}}{=} T_p\Rg T_p = G_1+ g^{-1} G_2 . 
\]
The assumption of the lemma implies $\tilde {\Rg}$ is invertible in 
$T_p \C^N$ and so is $A_3$ and 
$A_3^{-1}= \tilde{\Rg}^{-1}+ O(\lam^2g^2)$, which we substitute for 
for $A_3^{-1}$ in \refeq(Step3-3).  Then    
\begin{multline}  \lbeq(Step3-3a)
\refeq(Step3-3) = -N^{-1}g^{-1}\lam^{-2} (A+S)^{-1}S(A_1+T)^{-1}T
(A_2+T_p)^{-1} \\
\times  
T_p (\tilde{\Rg}^{-1}+ O(\lam^2g^2))T_p(A_2+T_p)^{-1}T(A_1+T)^{-1}S(A+S)^{-1}. 
\end{multline}
Then, we replace functions on each side of 
$(\tilde{\Rg}^{-1}+ O(\lam^2g^2))$ 
by the corresponding functions of \reflm(4-3) (2). 
The term $O(\lam^2g^2)$ produces 
$T_p O(g) T_p + O(\lam^2 g^3) = g S M(\lam)S + M(\lam)$  and 
$\tilde{\Rg}^{-1}$ does 
\[
-Ng^{-1}\lam^{-2}T_p \tilde{\Rg}^{-1} T_p + 
g (SM(\lam)\tilde{\Rg}^{-1}T_p + T_p \tilde{\Rg}^{-1} M (\lam) S) + M(\lam)
\]
Combining these results with \refeq(L2), we obtain \refeq(Gam-S3).  
\edpf

\newpage
\paragraph{Step 4.} Next we assume in addition to those of Step 3 
that $G_1$ is singular in $T_p \C^N$ but 
$G_1\not=0$. We let $T_e$ be the projection in 
$T_p \C^N$ onto $\Ker_{T_p\C^N}\,G_1$. 
We recall from \cite{CMY} that, then $T_e G_2 T_e $ is 
necessarily non-singular in $T_e \C^N$. 
In the following lemma we write $T_e^\perp$ for $T_p \ominus T_e$.
and we define the operator $\Bg$ on $T_p\C^N$ and $D_0$ on 
$(T_p\ominus T_e) \C^N$ by 
\begin{gather}
\Bg= T_e^\perp G_2 T_e (T_e G_2 T_e )^{-1} T_e, \\
D_0=(T_e^\perp G_1 T_e^\perp + g^{-1}
(T_e^\perp G_2 T_e^\perp - T_e^\perp G_2 T_e
(T_e G_2 T_e)^{-1}T_e G_2 T_e^\perp)^{-1}.
\end{gather}
By the remark above, $\Bg$ is well defined and, since 
$T_e^\perp G_1 T_e^\perp$ is clearly invertible in $(T_p \ominus T_e)\C^N$, 
$D_0$ is also well defined for small $\lam>0$.

\bglm \lblm(Step-4) Let $G_1$ and $T_e$ be as above.  
Then, $\Ga(\lam)^{-1}$ has the following expression as $\lam \to 0$:
\begin{multline}  \lbeq(Step-4)
- N^{-1}\lam^{-2} T_e  (T_eG T_e)^{-1} T_e 
-N^{-1}g^{-1}\lam^{-2}T_p (T_e^\perp - \Bg^\ast)T_e^\perp D_0 
T_e^\perp (T_e^\perp - \Bg)T_p \\
+ g^3 T_p M T_p + g^2 S M S + 
g S M + g M S + M.
\end{multline} 
\edlm 

For the proof we use the following well known formula from linear algebra.  

\bglm \lblm(FS) Suppose $a_{11}$ and $a_{22}$ are closed 
and $a_{12}$ and $a_{21}$ are bounded operators. 
Suppose that $a_{22}^{-1}$ exists. Then $A^{-1}$ exists if and only if 
$d= (a_{11}- a_{12}a_{22}^{-1}a_{21})^{-1}$ exists. In this case we have 
\bqn \lbeq(FS-formula)
A^{-1} = \begin{pmatrix}  d & -d a_{12} a_{22}^{-1} \\
-a_{22}^{-1}a_{21} d & a_{22}^{-1}a_{21}d a_{12} a_{22}^{-1} + a_{22}^{-1}
\end{pmatrix}.
\eqn 
\edlm

\bgpf 
We may repeat the argument of Step 3 upto \refeq(A3def) 
and notice that, as long as $A_3^{-1}$ exists in $T_p \C^N$, we have 
$\Ga(\lam)^{-1}= \refeq(Step3-2) + \refeq(Step3-3)$
and that $\refeq(Step3-2)= g S M(\lam) S+ M(\lam)$ (see \refeq(L2)).
Thus, we have only to study \refeq(Step3-3). 
We study $A_3^{-1}$ in $T_p\C^N$ by using \reflm(FS). 
We write $A_3$ in the block matrix in the 
direct decompotision $T_p\C^N= (T_p \ominus T_e)\C^N \oplus T_e \C_N $:
With obvious abusing notation 
\bqn  \lbeq(mat-1)
A_3 = 
\begin{pmatrix}  T_e^\perp G_1 T_e^\perp + g^{-1} 
T_e^\perp G_2 T_e^\perp  &  g^{-1} T_e^\perp G_2 T_e  \\
g^{-1} T_e G_2 T_e^\perp   & g^{-1} T_e G_2 T_e  \\
\end{pmatrix} + O(g^2 \lam^2)
\eqn 
Then, $a_{22}= g^{-1} T_e G_2 T_e$ is invertible in $T_e \C^N$ 
as mentioned above;  
\begin{align*}
& a_{11}- a_{12}a_{22}^{-1}a_{21}
= T_e^\perp G_1 T_e^\perp + g^{-1}(T_e^\perp G_2 T_e^\perp 
- T_e^\perp G_2 T_e
(T_e G_2 T_e)^{-1}T_e G_2 T_e^\perp)
\end{align*}
is also invertible for small $\lam>0$ 
in $T_p \C^N \ominus T_e\C^N$ because 
$\Ker\,G_1 \cap T_e^\perp=\{0\}$ by the definition of $T_e$. 
Then, \reflm(FS) yields that the matrix in the right of \refeq(mat-1) 
is invertible in $T_p\C^N$ and is equal to 
\bqn 
\begin{pmatrix} D_0 & - D_0 \Bg \\ \Bg^\ast D_0 & 
- \Bg^\ast D_0 \Bg + g (T_e G_2 T_e)^{-1} \end{pmatrix},
\eqn 
whichis of order $O(g)$ as $\lam \to 0$. Then, the standard 
perturbation theory implies 
\begin{align} 
A_3^{-1}& = 
\begin{pmatrix} D_0 & - D_0 \Bg \\ \Bg^\ast D_0 & 
\Bg^\ast D_0 \Bg+g (T_e G_2 T_e)^{-1} \end{pmatrix}
+ T_p O(g^4\lam^2)T_p  \notag \\
& = g (T_e G_2 T_e)^{-1} + (T_e^\perp-\Bg^\ast )D_0 (T_e^\perp + \Bg)
+ T_p O(g^4\lam^2)T_p .
\lbeq(A3inv)
\end{align} 
We  substitute \refeq(A3inv) for $A_3^{-1}$ in \refeq(Step3-3) 
and denote by $\Ga_1(\lam)$ and $\Ga_2(\lam)$ the functions 
produced respectively by $g (T_e G_2 T_e)^{-1}$ and  by the other two terms. 
Then, by virtue of \reflm(4-3) (2), $\Ga_1(\lam)$ 
is equal to 
\begin{align} 
& -N^{-1}\lam^{-2} 
(1 + \lam^2 g^2 S M  + \lam^2 g M )T_e(T_e G_2 T_e)^{-1}
T_e(1 + \lam^2 g^2 M S + \lam^2 g M).
\notag \\
& = -N^{-1}\lam^{-2}T_e(T_e G_2 T_e)^{-1}T_e  \lbeq(principal) \\
& \hspace{1cm} + g^2(SMT_e+ T_e MS)+ 
g (MT_e + T_e M) + M.  \lbeq(principal-rest)
\end{align}
Likewise, denoting $ \Cg=(T_e^\perp-\Bg^\ast )D_0 (T_e^\perp-\Bg)$, 
$\Ga_2(\lam)$ is equal to 
\begin{align} 
& -N^{-1}\lam^{-2}g^{-1} 
(1 + \lam^2 g^2 S M  + \lam^2 g M )T_p 
(\Cg + O(\lam^2 g^4))
T_p(1 + \lam^2 g^2 M S + \lam^2 g M) 
\notag \\
& = -N^{-1}\lam^{-2}g^{-1}T_p \Cg T_p + g^3 T_p M T_p 
+ g(SMT_p+ T_p MS) + M.  \lbeq(sub)
\end{align}
Combining \refeq(principal), \refeq(sub) with the remark stated 
at the beginning, we obtain \refeq(Step-4) 
and conclude the proof. 
\edpf 

\paragraph{Step 5.} We finally assume $G_1=0$ the opposite of 
of \reflm(Step-4). Then, $T_p=T_e$ and $H_{\a,Y}$ has no $p$-wave resonances.   

\bglm Suppose $G_1=0$. Then,  $T_e=T_p$ and 
\begin{multline} \lbeq(5-1)
\Ga(\lam)^{-1}= -N^{-1}\lam^{-2}T_e G_2^{-1} T_e + g^3 T_p M(\lam) T_p  \\
+ g^2 (T_e M(\lam) S + S M(\lam)T_e)+ 
g (T_e M(\lam)+ M(\lam)T_e) + M(\lam) .
\end{multline} 
\edlm 

\bgpf In this case we still have the expression 
$\Ga(\lam)^{-1}= \refeq(Step3-2) + \refeq(Step3-3)$ 
and \refeq(Step3-2) satisfies the estimate \refeq(L2). 
As was remarked previously  $G_2$ is non-singular in $T_p \C^N$. 
It follows from \refeq(A3def) that in $T_p \C^N$, 
\bqn \lbeq(a3-1)
A_3^{-1} = 
(g^{-1}G_2+ T_p O(\lam^2 g^2)T_p)^{-1}
= g G_2^{-1} + O(\lam^2 g^4).
\eqn 
We substitute \refeq(a3-1) in \refeq(Step3-3) and apply \reflm(4-3) (2). 
Then  $\refeq(Step3-3)$ becomes  
\begin{align*}
& - N^{-1}g^{-1}\lam^{-2}(T_p + \lam^2 g^2 S M + \lam^2 g M) 
T_p \\
& \hspace{2cm}
\times 
(g G_2^{-1} + O(\lam^2 g^4))T_p (T_p + \lam^2 g^2 MS + \lam^2 g M) \\
& =  - N^{-1}\lam^{-2}(T_p + \lam^2 g^2 S M+ \lam^2 g M)G_2^{-1}
(T_p + \lam^2 g^2 MS+ \lam^2 g M) + g^3 T_p M T_p \\
& = - N^{-1}\lam^{-2}T_p G_2^{-1}T_p 
-N^{-1}T_p G_2^{-1}(g^2 MS + g M) \\
& \hspace{2cm} - N^{-1}(g^2 SM + gM) G_2 T_p + M(\lam) 
+ g^3 T_p M T_p. 
\end{align*}
As $T_p= T_e$, this implies \refeq(5-1). 
\edpf

\subsection{Fourier transform of a logarithmic function}

In the following section need a pointwise estimate on  
the Fourier transform 
\bqn \lbeq(pgF)
F(x){=}
\frac1{2\pi}\int_{\R^2} \frac{e^{ipx }\chi_{\leq \ep}(|p|){dp}}{|p| g(|p|)}. 
\eqn 
It is obvious from Hausdorff-Young's inequality 
$F \in L^q(\R^2)$ for $2\leq q \leq \infty$. 
The following estimate must be well 
known and we give a proof for reader's convenience. 

\bglm \lblm(plogp)
Let $\chi_{\leq \ep}$ be defined by \refeq(chiep-def). Then for   
$0<\ep\leq 1$ there exists a constant $C_\ep >0$ such that 
\bqn \lbeq(plogp)
|F(x)|\leq \frac{C_\ep}{\ax \log(2+|x|)}
\eqn 
\edlm 
\bgpf Since $F(x)$ is evidently a smooth function, it suffices to show 
\refeq(plogp) when $|x|$ is sufficiently large and we assume  
$|x|>(100e^{10\c}+4\pi/\ep)$, $\c= 0.577\dots$ being Euler's constant.  Since 
$g(r)\geq (\log |x|)/200\pi +1/8$ for $0<r<2\pi/|x|$, we have  
\[
\frac1{2\pi}\int_{\R^2} \frac{e^{i\xi x }
\chi_{\leq 2\pi/|x|}(|\xi |)\chi_{\leq \ep}(|\xi |){d\xi }}{|\xi | g(|\xi |)} 
\absleq 
\int_0^{2\pi/|x|} \frac{dr}{|g(r)|}\leq 
\frac{C}{\ax \log(2+|x|)} 
\]
for a constant $C>0$. Thus, it suffices to prove the lemma 
after inserting $\chi_{\geq 2\pi/|x|}(|\xi |)$ in the integrand 
of \refeq(plogp). We denote the function thus obtained by $\tilde{F}(x)$.  
The Bessel function satisfies (see e.g. \cite{Stein}, page 338):
\[
J_0(r)= \frac1{\pi}\int_{0}^{\pi} e^{ir\cos\th}d\th 
= (2/\pi)^{1/2}r^{-1/2}\cos(r-\pi/4)+ O(r^{-3/2}),  \quad r \to \infty.
\]
Let $\chi_{[\ep_1,\ep_2]}(r) 
= \chi_{\geq \ep_1}(r)\chi_{\leq \ep_2}(r)$. Then, after a change variables 
\begin{align}
\tilde{F}(x)&= 
\frac1{\sqrt{2\pi}}\int_0^\infty \frac{\chi_{[2\pi/|x|,\ep]}(r)}{g(r)}
\Big(
\frac{\cos(|x|r-\pi/4)}{\sqrt{|x|r}}+ O((|x|r)^{-3/2})\Big)dr \notag \\ 
& = \frac1{\sqrt{2\pi}|x|}\int_{0}^\infty 
\frac{\chi_{[2\pi,\ep|x|]}}{g(r/|x|)}\Big(\frac{\cos(r-\pi/4)}{\sqrt{r}}
+ O(r^{-3/2})\Big)dr . \lbeq(tilF)
\end{align}
Write $\tilde{\chi}$ for $\chi_{[2\pi,\ep|x|]}$. 
Denote the integral produced by  $O(r^{-3/2})$ in \refeq(tilF) by $F_1(x)$. 
Since $|g(r)|\geq 1/4$ for any $r>0$ and 
$|g(r/|x|)|\geq C \log |x|$ for $2\pi \leq r<{\sqrt{x}}$  as 
$|x|>100e^{10\c}$ then, we have 
\begin{align*}
|F_1(x)| & \leq   
\frac1{|x|}\left(
\int_0^{\sqrt{x}}+ \int_{\sqrt{x}}^{\ep |x|} \right)
\frac{\tilde{\chi}(r)} {|g(r/|x|)|}\la r \ra ^{-3/2}dr \\
&\leq 
\frac{2}{|x|}\int_0^{\sqrt{x}}\frac1 {g(r/|x|)}\la r \ra ^{-3/2}dr
+ \frac{C}{|x|}\int_{\sqrt{x}}^{\ep |x|} \la r \ra ^{-3/2}dr 
\leq \frac{C}{\ax \log(2+ |x|)}. 
\end{align*}.
Since $\cos(s+\pi)= -\cos s$ and $2\pi<r<\ep|x|$, we have 
\begin{align*}
& 
\frac1{|x|}\int_0^\infty \frac{\tilde{\chi}(r)}{g (r/|x|)}
\frac{\cos(r-\tfrac{\pi}{4})}{r^{1/2}} dr 
= - \frac1{|x|}\int_0^\infty 
\frac{\tilde{\chi}(r+\pi)}{g ((r+\pi)/|x|)} 
\frac{\cos(r-\tfrac{\pi}{4})}{(r+\pi)^{1/2}}dr \\
& = \frac1{2|x|}\int_0^\infty \cos(r-\tfrac{\pi}{4}) 
\left(\frac{\tilde{\chi}(r)}{g(r/|x|)} r^{-1/2}- 
\frac{\tilde{\chi}(r+\pi)}{g((r+\pi)/|x|)} (r+\pi)^{-1/2}\right) dr 
\end{align*}
The function inside $(\cdots )$ is $K_1+ K_2+ K_3$ where  
\begin{align*}
K_1(r,x) & = \frac{\tilde{\chi}(r)-\tilde{\chi}(r+\pi)}{g (r/|x|)} r^{-1/2}. \\
K_2 (r,x) & = \tilde{\chi}(r+\pi)
\left(\frac1{g (r/|x|)} -\frac1{g((r+\pi)/|x|)}\right) 
r^{-1/2} . \\
K_3 (r,x) & = 
\frac{\tilde{\chi}(r+\pi)}{g((r+\pi)/|x|)}(r^{-1/2}-(r+\pi)^{-1/2}) .
\end{align*}
Since $\tilde{\chi}(r)-\tilde{\chi}(r+\pi)\not=0$ only on  
$[\pi,2\pi]$ and on $[\ep|x|-\pi, \ep|x|]$ and,  
$|g(r/|x|)|\geq c\log |x|$ on  $[\pi,2\pi]$ and 
$|g(r/|x|)|\geq C$ on $[\ep|x|-\pi, \ep|x|]$, it follows that 
\[
\frac1{2|x|}\int_0^\infty \cos(r-\tfrac{\pi}{4}) K_1(r,x)dr 
\absleq \frac{C}{|x|}\left( 
\int_{\pi}^{2\pi}\frac{dr}{\log |x|}dr + 
\int_{\ep|x|-\pi}^{\ep|x|}\frac{dr}{r^{1/2}}dr\right)  
\]
and this is bounded in modulus by $C\ax^{-1}(2+ \log |x|)^{-1}$ as 
desired. Note that $K_2(r,x)\not=0$ for $\pi <r <\ep |x|-\pi$ 
and we estimate 
\begin{align*}
&\frac1{g(r/|x|)} -\frac1{g((r+\pi)/|x|)} 
\absleq \frac{\log(r+\pi)-\log r}
{2\pi g(r/|x|) \cdot g((r+\pi)/|x|)} \\
&= \frac{\log (1+\frac{\pi}{r})}{2\pi g(r/|x|)\cdot g((r+\pi)/|x|)}
\leq \left\{
\br{cl} \displaystyle
\frac{C_\ep }{(\log |x|)^2 r}, & \ 2\pi<r<\sqrt{|x|}, \\[10pt]
\displaystyle \frac{C_\ep}{r}, & \ \sqrt{|x|}\leq r \leq \ep |x| .  
\er 
\right.
\end{align*}
It follows for large $|x|$ that 
\begin{align*}
& \frac1{2|x|}\int_0^\infty \cos(r-\tfrac{\pi}{4}) K_2(r,x)dr \\
& \absleq 
\frac{C}{|x|} \int_{2\pi}^{\sqrt{x}}
\frac{1}{(\log x)^2 r^\frac32}dr +\frac{C}{|x|} \int_{\sqrt{x}}^{\ep |x|}
\frac{1}{r^\frac32}dr 
\leq \frac{C_1}{|x|(\log |x|)^2}. 
\end{align*}
We estimate for $\pi \leq r \leq \ep |x|-\pi$ 
\[
K_3 (r,x) \absleq \frac{\pi\tilde{\chi}(r+\pi)}
{r^{1/2}(r+\pi)|g((r+\pi)/|x|)|}
\leq \frac{C}{r^{3/2}}
\]
and 
\[
\frac1{2|x|}\int_0^\infty \cos(r-\tfrac{\pi}{4}) K_3(r,x)dr
\absleq 
\frac1{2|x|}\int_{\pi}^{\ep|x|-\pi}\frac{dr}{r^{3/2}} 
\leq \frac{C}{\ax \log (2+|x|)}.
\]
Adding these up, we complete the proof. \edpf 

\section{Proof of \refthb(Main)} 

In this section we prove \refth(Main). 
By virtue of \refth(high) it suffices to prove the statements for 
the lower energy part $W_{\rm low}^{-} = W^{-}\chi_{\leq \ep}(|D|)$. 
By virtue of \refeq(Stat-2a), 
$W_{\rm low}^{-}= \chi_{\leq \ep}(|D|) u + \W_{\rm low}u$, where   
\bqn 
\W_{\rm low}u= 
\frac1{\pi{i}}\int_0^\infty \chi_{\leq \ep}(\lam) \lam \big\la 
\tGa(\lam) \hat{\Gg}_{\lam}(x), 
\int_{\R^2}(\hat{\Gg}_{\lam}(y)-\hat{\Gg}_{-\lam}(y))u(y) dy \big\ra_{\C^N} 
d\lam 
\lbeq(low-W)
\eqn 
and we study $\W_{\rm low}$. Recall $\tGa(\lam)  = \Ga(\lam)^{-1}$. 
Here and hereafter we omit the index $Y$ and write $\hat{\Gg}_{\lam}(x)$ for 
$\hat{\Gg}_{\lam,Y}(x)$  and etc. $\|u\|_p$ is the norm of $L^p(\R^2)$ 
and $\|u\|=\|u\|_p$. $\la u, v\ra$ 
is the coupling without complex conjugation. The inner product of 
$L^2$ will be denoted by $(u,v)$. Recall the space $\Dg_\ast$ is defined 
by \refeq(Dgast-def) and is dense in $L^p(\R^2)$ for any $1<p<\infty$.
In what follows  it is implicitly assumed that 
$u \in \Dg_\ast$.

\subsection{Proof of statement (1)} 
If $S\tD{S}$ is non-singular in $S\C^N$, then \reflm(Step1) implies that 
$\tGa(\lam)\chi_{\leq \ep}(\lam)$ for a small $\ep>0$ is 
a good multiplier. The product formula \refeq(prod) implies 
\[
(\W_{jk}\chi_{\leq \ep}(|D|)u)(x)= (K \circ \tGa_{jk}(|D|)\chi_{\leq \ep}(|D|))u(x) 
\]
and \reflm(prod) and \reflm(Mikhlin) imply statement (1). 
This has been proved already in \cite{CMY}.  
 
\subsection{Proof of statement (2)} 

Under the condition of statement (2), \reflm(Ga-pwave) is satisfied and 
we have \refeq(S2-extra) for $\lam $ in the support of the function
$\chi_{\leq \ep}$. We write $B$ for $(T\tD^2 T)^{-1}$ 
and substitute 
\refeq(S2-extra) for $\tGa(\lam)$ in \refeq(low-W). The  
term $M(\lam)$ produces a good operator as in the proof of statement (1) 
and we are left with 
\bqn \lbeq(low-Wa)
\frac{N}{i\pi}\int_0^\infty \chi_{\leq \ep}(\lam) 
\lam g 
\big\la TBT \hat{\Gg}_{\lam}(x) ,
\int_{\R^2}(\hat{\Gg}_{\lam}(y)-\hat{\Gg}_{-\lam}(y))u(y)dy \big\ra d\lam.  
\eqn
For \refeq(low-Wa), we may still apply the product decomposition 
of \reflm(prod), however, the multiplier 
$\chi_{\leq\ep} (\lam)g(\lam)$ is not bounded 
near $\lam=0$ and Mikhlin's theorem does not apply. To get around this  
we use the cancellation produced by the projector $T$ 
among the components of 
$\hat{\Gg}_{\lam}(y)-\hat{\Gg}_{-\lam}(y)$. 
Define  the vector function $\hat{I}(\lam)$ by 
\[
\hat{I}(\lam)= \frac{1}{i\pi}
\int_{\R^2}(\hat{\Gg}_{\lam}(y)-\hat{\Gg}_{-\lam}(y))u(y)dy
= \int_{{\mathbb S}^1} \begin{pmatrix} 
e^{iy_1\w\lam} \\
\vdots \\
e^{iy_N\w\lam} 
\end{pmatrix}\hat{u}(\lam\w)dw.
\]
Since $T=ST$, we may insert the orthogonal projection  $S$ 
in front of $I(\lam)$  without changing \refeq(low-Wa).
Since $NS=N-{\bf 1}\otimes {\bf 1}$, we then have  
\bqn \lbeq(NSpi)
NS \hat{I}u(\lam)= \int_{{\mathbb S}^1}
\sum_{k=1}^N 
\begin{pmatrix} 
e^{iy_1\w\lam}-e^{iy_k\w\lam}  \\
\vdots \\
e^{iy_N\w\lam} -e^{iy_k\w\lam} 
\end{pmatrix} \hat{u}(\lam\w)dw.
\eqn 
Recall that $R=\begin{pmatrix} R_1 \\ R_2 \end{pmatrix}$ 
and $R_l= R_l(D)= \Fg^{-1}(\xi_l/|\xi|)\Fg$, $l=1,2$ are 
Riesz transforms and $\la a, R u\ra= a_1 R_1 u + a_2 R_2 u$ 
for a vector $a\in \R^2$. Then, Taylor's formula implies that the $j$-th 
component of \refeq(NSpi) is equal to   
\begin{align}
& i \sum_{k=1}^N \int_{{\mathbb S}^1}\left(\int_0^1 
e^{i(\th {y_j}+ (1-\th)y_k)\w\lam} (y_j-y_k) \w\lam\hat{u}(\lam\w)d\th\right) 
d\w   \notag \\ 
&=i\lam \sum_{k=1}^N \int_0^1 \Fg \left(
\int_{{\mathbb S}^1}\la y_j-y_k, R \tau_{\th{y_j}+ (1-\th)y_k}u\ra )(\lam\w) d\w 
\right) d\th, \lbeq(UseR) 
\end{align}
where $\tau_y u(x)= u(x-y)$ is the translation by $y$. 
Define 
\bqn 
m(\lam) = \lam g(\lam)\chi_{\leq \ep} (\lam).      \lbeq(m-def)
\eqn 
Then, $m(|\xi|)$ is a good multiplier, $m(|D|)$ commutes 
with translations and $R$ and we have  
\bqn 
m(\lam)
\Fg \big(\la a, R \tau_{y}u\ra \big)(\lam\w)
= \Fg \big(\la a, R \tau_{y}m(|D|)u\ra \big)(\lam\w).
\eqn 
Thus, if we define 
$(V_{jk}u)(x) 
= \int_0^ 1 \la y_j-y_k, R \tau_{\th{y_j}+ (1-\th)y_k}m(|D|)u\ra d\th$ 
for $j,k=1, \dots, N$, then, $V_{jk}$ are evidently good operators 
and \refeq(DEFK) implies  
\begin{align}
\refeq(low-Wa)& = \int_0^\infty \chi_{\leq \ep}(\lam)\lam g(\lam)
\big\la TBT \hat{\Gg}_{\lam}(x) , NS  \hat{I}u(\lam) \big\ra 
d\lam. \notag \\
& = \sum_{jkl}(TBT)_{lj} 
\int_0^\infty \lam {\Gg}_{\lam}(x-y_l) \left(\int_{{\mathbb S}^1}
(\Fg V_{jk}u)(\lam\w)d\w \right)d\th \notag \\
& = \sum_{jkl} (TBT)_{jk} \tau_{y_l}(K\circ V_{jk}u)(x).  \lbeq(3-36a)
\end{align}
This  is a good operator and statement (2) is proved. 
\qed

\subsection{Proof of statement (3)}

We next assume that $T_p \Gg_1(Y) T_p = G_1$ is non-singular in $T_p\C^N$ 
and $\Ga(\lam)^{-1}$ satisfies \reflm(p-wave), 
We substitute \refeq(Gam-S3) for $\tGa(\lam)$ in 
\refeq(low-W), which produces three operators. The proof of 
statements (1) and (2) implies that $gS M(\lam) S$ and $M(\lam)$ produce  
good operators. Ignoring unimportant constant, we write the integral produced by 
$-N^{-1}g^{-1}\lam^{-2} T_p (G_1 + g^{-1}G_2)^{-1}T_p$ in the form  
\bqn \lbeq(low-W-vec) 
\W_{\rm low}u(x)=
\int_0^\infty \chi_{\leq \ep}(\lam)\lam^{-1} g(\lam)^{-1} 
\big\la T_p B_\ast T_p \hat{\Gg}_{\lam}(x) , 
\frac{NS}{i\pi}\hat{I}u(\lam) \big\ra_{\C^N} 
d\lam ,
\eqn 
where we wrote $ST_p = T_p$ for simplicity and $B_\ast$ for  
\[
B_\ast(\lam)=(G_1 + g^{-1}G_2)^{-1}.
\] 
Remark that we have inserted $S$ in front of $\hat{I}u(\lam)$ 
which is allowed by the presence of $T_p$. Notice  
the presence of the strong singularities $\lam^{-1} g(\lam)^{-1}$. 

\subsubsection{Decomposition into good part and bad parts.} 
By further expanding the exponential functions, we 
decompose $NS\hat{I}u(\lam)$ into ``good'' and ``bad'' 
parts $\hat g(\lam)$ and $\hat{b}(\lam)$ as follows:
\begin{gather}
(NS\hat{I} u)(\lam) = \hat{g}(\lam) + \hat{b}(\lam)
= \begin{pmatrix} g_1(\lam) \\ \vdots \\ g_N(\lam)  \end{pmatrix}
+  \begin{pmatrix} b_1(\lam) \\ \vdots \\ b_N(\lam)   \end{pmatrix}, 
\lbeq(good-bad) \\
g_j(\lam)=\sum_{k=1}^N g_{jk}(\lam), \ \  
b_j(\lam)= \sum_{k=1}^N b_{jk}(\lam) \ \ j=1, \dots, N  \\
g_{jk}(\lam)\stackrel{\mathrm{def}}{=}  i 
\int_{{\mathbb S}^1}\left(
\int_0^1 
(e^{i(\th {y_j}+ (1-\th)y_k)\w\lam} -1)
 (y_j-y_k) \w\lam\hat{u}(\lam\w)d\th\right) 
d\w, \lbeq(good) \\
b_{jk}(\lam)\stackrel{\mathrm{def}}{=}  i 
\int_{{\mathbb S}^1} (y_j-y_k) \w\lam\hat{u}(\lam\w)d\w \,. \lbeq(bad)
\end{gather}
Then, substituting \refeq(good-bad) for $(NS\hat{I} u)(\lam)$ 
in \refeq(low-W-vec), we obtain 
\bqn \lbeq(gb-decomp)
\W_{\rm low}u = \W_{\rm low, g}u + \W_{\rm low,b}u 
\eqn 
where definition of $\W_{\rm low, g}u$ and $\W_{\rm low,b}u$ should be obvious. 

\subsubsection{Good part produces a good operator}
\bglm \lblm(good)
For $1<p<\infty$, there exists a constant $C_p>0$ such that 
\bqn \lbeq(good-part)
\|\W_{\rm low, g}u\|_p \leq C_p \|u\|_p, \quad u \in \Dg_\ast. 
\eqn 
\edlm 
\bgpf Taylor's formula implies    
\[
e^{i(\th {y_j}+ (1-\th)y_k)\w\lam} -1= 
i\lam \la \th {y_j}+ (1-\th)y_k, \w \ra \cdot 
\int_0^1 
e^{i\mu(\th {y_j}+ (1-\th)y_k)\w\lam}d\mu, 
\]
which produces an extra factor in \refeq(good) and 
$g_{jk}(\lam)$ becomes the integral $d\th{d\mu}$ over 
$(0,1) \times (0,1)$ of 
\begin{multline*}
-\lam^2  
\int_{\mathbb{S}^1} 
e^{i\mu(\th {y_j}+ (1-\th)y_k)\w\lam} 
(\la y_k, \w\ra \la y_j-y_k, \w\ra +\th\la y_j-y_k,\w\ra^2) \hat{u}(\lam\w) 
d\w  \\
=-\lam^2 
\int_{\mathbb{S}^1} \Fg(
\t_{\mu(\th {y_j}+ (1-\th)y_k)} 
(\la y_k, R\ra \la y_j-y_k, R\ra +\th\la y_j-y_k,R\ra^2){u})(\lam\w) d\w.
\end{multline*}
Since $\{\tau_y : y \in \R^2\}$ is uniformly bounded in $\Bb(L^p)$, the operator 
\[
W_{jk}u(x)\stackrel{\rm def}{=} \iint_{[0,1]^2} 
\t_{\mu(\th {y_j}+ (1-\th)y_k)} 
(\la y_k, R\ra \la y_j-y_k, R\ra +\th\la y_j-y_k,R\ra^2){u})(x)d\th{d\mu}.
\] 
is a good operator. Define $m(\lam) = g^{-1}(\lam)\chi_{\leq \ep}(\lam)$. 
$m(\lam)$ is a good multiplier and we can express $\W_{\rm low,g}u(x)$ 
is the form  
\[
\sum_{j,k,l=1}^N  
\int_0^\infty \lam  (T_p B_\ast(\lam) T_p)_{jk}\Gg_{\lam}(x-y_k)
\left( 
\int_{{\mathbb S}^1}\Fg(W_{jl}m(|D|)u)(\lam\w) d\w
\right)
d\lam .
\]
It follows by virtue of the definition \refeq(DEFK) of $K$ that 
\bqn \lbeq(Wgood)
\W_{\rm low,g}u(x) = \pi^2 \sum_{j,k,l=1}^N \tau_{y_k} 
(K\circ W_{jl}m(|D|)(T_p B_\ast(|D|) T_p)_{jk} u)(x). 
\eqn 
Since $B_\ast(\lam)$ is a good multiplier under the assumption, 
\reflm(prod) implies the lemma.  
\edpf 

\subsubsection{Decomposition of the bad part}
We decompose 
$\W_{\rm low,b}u(x)$ into the low and high energy parts: 
\begin{align}
\W_{\rm low,b}u(x)& = \int_0^\infty 
\lam^{-1} \chi_{\leq \ep}(\lam) g(\lam)^{-1}
\big\la T_p B_\ast(\lam) T_p 
\hat{\Gg}_{\lam}(x), \hat{b}(\lam)\big\ra d\lam 
\lbeq(x-decomp0) \\
& = \chi_{\geq 2\ep}(|D|)\W_{\rm low,b}u(x) +  
\chi_{\leq 2\ep}(|D|)\W_{\rm low,b}u(x)  \lbeq(x-decomp). 
\end{align}
Note that supports of $\chi_{\geq 2\ep}$ and $\chi_{\leq \ep}$ do not 
intersect. 

\bglm \lblm(Bad-B) 
For any $\ep>0$, $\chi_{\geq 2\ep}(|D|)\W_{\rm low,b}$ 
is bounded from $L^p(\R^2)$ to itself for $1<p\leq 2$.  
\edlm 
\bgpf Denote $\tilde{B}_{jk}=(T_p B_\ast T_p)_{jk}$ and 
express $\chi_{\geq 2\ep}(|D|)\W_{\rm low,b}u(x)$ as the sum over 
$j,k=1, \dots, N$ of  
\bqn \lbeq(Xjk)
X_{jk}u(x)\stackrel{\rm def}{=}\int_0^\infty 
\lam^{-1} \chi_{\leq \ep}(\lam) g(\lam)^{-1} \tilde{B}_{jk}(\lam)
\t_{y_j}\chi_{\geq 2\ep}(|D|){\Gg}_{\lam}(x)b_k(\lam) d\lam .
\eqn 
Define $\m(\xi)= \chi_{\geq 2\ep}(|\xi|)|\xi|^{-2}$. $\m$ is a good multiplier. 
By applying the inverse Fourier tansform to 
$\chi_{\geq 2\ep}(\xi)(\xi^2-\lam^2)^{-1}= 
\mu(\xi) + \lam^2 \mu(\xi) (\xi^2-\lam^2)^{-1}$ we have  
\bqn 
\chi_{\geq 2\ep}(|D|){\Gg}_{\lam}(x)= \hat{\m}(x) + 
\m(|D|) \lam^2 {\Gg}_{\lam}(x). \lbeq(chigeq)
\eqn 
We substitute \refeq(chigeq) for the 
$\chi_{\geq 2\ep}(|D|){\Gg}_{\lam}(x)$ in \refeq(Xjk). The second summand  
$\m(|D|) \lam^2 {\Gg}_{\lam}(x)$ cancels the singularity $\lam^{-1}$ 
and produces 
\bqn \lbeq(mDint)
\sum_{l=1}^N \m(D) 
\int_0^\infty \lam
{\Gg}_{\lam}(x-y_j) \left(\int_{{\mathbb S}^1} 
\la y_k-y_l, \w \ra \r_{jk}(\lam) 
\hat{u}(\lam\w)d\w \right) d\lam
\eqn 
where $\r_{jk}(\lam)= \lam\chi_{\leq \ep}(\lam)g(\lam)^{-1}\tilde{B}_{jk}(\lam)$ 
is obviously a good multiplier.
By using $K$ of \refeq(DEFK), we may express \refeq(mDint) in the form   
\[
\sum_{l=1}^N\m(D)\t_{y_j} K \circ (\la y_k -y_l, R \ra \r_{jk}(|D|)) u  
\]
and \reflm(prod) implies that this is a good operator.  

The first summand $\hat{\m}(x)$ produces     
\bqn \lbeq(Int-1)
\sum_{l=1}^N  \int_0^\infty \chi_{\leq \ep}(\lam) g(\lam)^{-1}
\hat{\m}(x-y_j) \tilde{B}_{jk}(\lam)\big(\int_{{\mathbb S}^1} 
i \la y_k-y_l, \w\ra \hat{u}(\lam\w)d\w \big) d\lam .
\eqn 
Here $\hat{\m}(x)$ is $\lam$-independent and in the polar coordinates 
$\xi = \lam \w$, $\lam>0$ and $\w \in {\mathbb S}^1$, 
$d\xi=\lam d\lam d\w $. Thus, by using the Parseval formula, we may express   
\bqn \lbeq(N-2)
\refeq(Int-1) = \sum_{l=1}^N  i \hat{\m}(x-y_j) 
\la \Fg \tilde \r_{jk}, \la y_k-y_l,R \ra u\ra_{L^2},
\eqn 
where $\tilde\rho_{jk}(\xi)= |\xi|^{-1}\chi_{\leq \ep} (|\xi|)
g(|\xi|)^{-1}\tilde{B}_{jk}(|\xi|)$. It is evident that  
$\tilde\rho_{jk}\in L^p(\R^2)$ for $1<p \leq 2$ and, 
we have $\hat{\mu} \in L^p(\R^2)$ for $1\leq p<\infty$ because 
$\hat\m(x) = N_0(x) - (N_0 \ast \widehat{\chi_{\leq 2\ep}})(x)$ 
implies 
\bqn \lbeq(hatm)
|\hat{\mu}(x)| \leq C_N |\log |x|| \ax^{-N}, \quad N=0,1, \dots.
\eqn 
Thus H\"older's  and Hausdorff-Young's inequalities imply 
for $1<p\leq 2$ and its dual exponent $q=(p-1)/p$ that  
 \begin{multline*}
\|\hat{\m}(x-y_j) 
\la \Fg \tilde \r_{jk}, \la y_k-y_l,R \ra u\ra\|_p \\
\leq \|\hat{\mu}\|_p \|\Fg \tilde \r_{jk}\|_{L^q}
\|\la y_k-y_l,R \ra u\ra\|_p 
\leq C \|\hat{\mu}\|_p \|\tilde \r_{jk}\|_{p}\la u\ra\|_p 
\end{multline*}
and \refeq(Int-1) is bounded in $L^p(\R^2)$ for $1\leq p\leq 2$.  
\end{proof} 

We next show that 
$\chi_{\geq 2\ep}(|D|)\W_{\rm low,b}$ 
is unbounded in $L^p(\R^2)$ for $2<p<\infty$. 

\bglm \lblm(Bad-C) Suppose $\ep>0$ is sufficiently small. 
Then, $\chi_{\geq 2\ep}(|D|)\W_{\rm low,b}$ 
is unbounded from $L^p(\R^2)$ to itself for any $2<p<\infty$.  
\edlm 
\bgpf By virtue of the proof of previous \reflm(Bad-B), it suffices to 
show that the sum over $j,k=1, \dots, N$ of \refeq(Int-1)  or \refeq(N-2) 
is unbounded in $L^p$ for $p>2$. Introduce the notation: 
$\r(\lam) = \chi_{\leq \ep}(\lam) g(\lam)^{-1}|{\lam}|^{-1}$ and 
\bqn \lbeq(notat-hat)
\hat{\m}_Y(x) = \begin{pmatrix}\hat{\m}(x-y_1) 
\\ \vdots \\ \hat{\m}(x-y_N) \end{pmatrix}, \quad 
R_Y(\w) = 
\begin{pmatrix}\la y_1, \w \ra \\ \vdots \\ \la y_N, \w \ra \end{pmatrix}, 
\eqn 
Because of the presence of $T_p$ in $\tilde{B}= T_p {B_\ast}(\lam) T_p$ 
which annihilates $P$, we have again with vector notation and 
$\w=\xi/|\xi|$ that  
\begin{align}
& \sum_{jkl}\refeq(Int-1) = 
\int_0^\infty \r(\lam) 
\la (T_p {B_\ast}(\lam)T_p) \hat{\m}_Y(x), \hat{b}(\lam)\ra_{\C^N} d\lam \notag \\
& = \Big\la T_p \hat{\m}_Y(x), 
\int_0^\infty \r(\lam)\lam^{-1}
\left(\int_{{\mathbb S}^1} {B_\ast(\lam)} T_p R_Y(\w) \hat{u}(\lam\w)d\w \right) 
d\lam \Big\ra 
\notag \\
& = \Big\la T_p \hat{\m}_Y(x), 
\int_{\R^2} {B_\ast(|\xi|)} 
T_p R_Y(\w) \r(|\xi|)\hat{u}(\xi)d\xi \Big\ra , \notag  \\
& = \Big\la T_p \hat{\m}_Y(x), 
\int_{\R^2} \Fg(B_\ast(|\xi|)T_p R_Y(\w) \r(|\xi|))(y)u(y) dy \Big\ra 
\lbeq(Bad-Ia)
\end{align}
where we used Parseval identity in the last step. 
Take an orthonormal basis $\{\eb_1, \dots, \eb_n \}$ of $T_p \C^N$ 
such that $T_p \Gg_1(Y) T_p \eb_j 
= \k_j \eb_j$, $j=1, \dots, n$, $n ={\rm rank\ }T_p$ (we shall see 
$\k_j>0$ shortly). Then, \refeq(Bad-Ia) is equal to 
\bqn 
\sum_{j=1}^n \la \eb_j, \hat{\m}_Y(x) \ra \int_{\R^2}
\Fg (\la \eb_j, B_\ast R_Y(\w) \ra \r(|\xi|))(y)u(y) dy 
\lbeq(Bad-I).  
\eqn 
We note that,  if $y_j\not=y_k$ for $j\not=k$, 
any non-trivial linear combination  of 
$\hat{\m}(x-y_1), \dots, \hat{\m}(x-y_N)$ does not vanish because that 
\[
\Fg(c_1 \hat{\m}(x-y_1)+  \cdots + c_N \hat{\m}(x-y_N))(\xi)
= |\xi|^{-2}\chi_{\geq 2\ep}(\xi) \sum_{j=1}^N c_j e^{iy_j \xi}= 0 
\]
implies 
$\sum_{j=1}^N c_j e^{iy_j \xi}=0$ for $|\xi|\geq 2\ep$ and hence 
$c_1= \dots = c_N =0$. It follows that 
$\sum_{j=1}^n c_j \la \eb_j, \hat{\m}_Y(x)\ra =0$ 
implies $\sum_{j=1}^n c_j \eb_j=0$, 
hence $c_1= \dots= c_n=0$, viz. 
$\{\la \eb_j, \hat{\m}_Y(x) \ra \colon j=1, \dots, n\}$ is linearly independent. 
Then, Hahn-Banach theorem implies for any $j=1, \dots, n$, there exists 
$ z_j \in L^q$, $q=p/p-1$ such that 
\[
\int_{\R^2}z_j(x) \la \eb_k, \hat{\m}_Y(x) \ra dx=\d_{jk}, \quad k=1, \dots, n.
\] 
It follows that if $\sum_{jkl}\refeq(Int-1)$ is bounded in $L^p(\R^2)$ then 
\bqn \lbeq(by2)
u \mapsto \int_{\R^2} 
\Fg (\la \eb_j, B_\ast R_Y(\w) \ra \r(|\xi|))(y)u(y) dy, \quad j=1, \dots, n 
\eqn 
must be bounded linear functionals on $L^p(\R^2)$ or, by virtue of the Riesz 
representation theorem,  
$\Fg (\la \eb_j, B_\ast R_Y(\w) \ra \r(|\xi|)) \in L^q(\R^2)$, 
$j=1, \dots, n$, 
$q$ being the dual exponent of $p$, hence $1<q<2$. Then, 
by virtue of Hausdorff-Young's inequality we must have 
$B_\ast(|\xi|) R_Y(\w) \r(|\xi|) \in L^p(\R^2)$, 
which is equivalent to 
$T_p R_Y(\w)\r(|\xi|)\in L^p(\R^2)$, or 
$\la \eb_j, R_Y(\w) \ra \r(|\xi|)\in L^p(\R^2)$, $j=1, \dots,n $ 
because $\la \eb_j, B_\ast(|\xi|) \eb_k\ra = \k_j \d_{jk} + O(g^{-1})$, 
$\k_j>0$, if $\ep>0$ is sufficiently small. 
However, if $e^{(j)}_l$ is the $l$-th component of $\eb_j$, we have 
\bqn \lbeq(by1)
\k_j = \la \eb_j, T_p \Gg_1(Y) T_p \eb_j\ra = -\frac12 \sum_{k,l=1}^N 
|y_k-y_l|^2 e^{(j)}_k e^{(j)}_l = \big(\sum_{l=1}^N e^{(j)}_l y_l \big)^2 
\eqn 
and for every $j=1, \dots, n$, 
\bqn \lbeq(by)
\la \eb_j ,\yb \ra 
= \sum_{l=1}^N e^{(j)}_l y_l \not=0, \quad \yb= 
\begin{pmatrix}y_1 \\ \vdots \\ y_N \end{pmatrix} . 
\eqn 
Thus, $\la \eb_j, R_Y(\w) \ra \r(|\xi|)\not\in L^p(\R^2)$, $2<p<\infty$ 
for any $1\leq j \leq n$ because  
\[
\int_{|\xi|<\ep} \frac{|\la \ab, \w \ra|^p}{|\xi|^p|g(|\xi|)|^p}d\xi = \infty
\]
for any $\ep>0$ and any $\ab \in \R^2 \setminus\{0\}$.  
This complete the proof.
\edpf 

The proof of statement (3) will be finished if we have proven the 
following lemma. 

\bglm \lblm(2-final) Operator $\chi_{\leq 2\ep}(|D|)\W_{\rm low,b}$ 
is bounded in $L^p(\R^2)$ if $1<p< 2$. 
\edlm 
\bgpf The proof uses the fact that 
$\tilde B(\lam) = T_p B_\ast(\lam) T_p$ has the factor $T_p$ also on the 
right. $\chi_{\leq 2\ep}(|D|)\W_{\rm low,b}(x)$ 
may be expressed in the form   
\begin{align} \lbeq(6-16)
& i \int_0^\infty \int_{{\mathbb S}^1} \chi_{\leq \ep}
(\lam) g(\lam)^{-1} 
\big\la \tilde{B} \chi_{\leq 2\ep}(|D|)\hat{\Gg}_{\lam} (x), 
\la\yb, \w\ra \big\ra
\hat{u}(\lam\w) 
d\lam d\w  \\
& = \sum_{jk}\int_{\R^4}
\frac{\chi_{\leq {2\ep}}(\xi)\chi_{\leq\ep}(|\eta|)\tilde{B}_{jk}(|\eta|)}
{(\xi^2-\eta^2-i0)|\eta|g(|\eta|)} 
e^{i(x-y_j)\xi}\la y_k, \hat{\eta}\ra \hat{u}(\eta) d\eta d\xi . \lbeq(6-16a)
\end{align}
Recall that we are assuming $u \in \Dg_\ast$ 
and $(\xi^2-\eta^2-i0)^{-1}$ in \refeq(6-16a) is well defined as 
limit $\k \downarrow  0$ of $\lim (\xi^2-\eta^2-i\k)^{-1}$.  
Hereafter in the proof we write  
$\r_{jk}(\lam)=\chi_{\leq\ep}(\lam)\tilde{B}_{jk}(\lam)g(\lam)^{-1}$. 
$\r_{jk}$ is a good multiplier. 

Since $\tilde{B}= T_p B_\ast T_p$ and $T_p$ annihilates 
${\bf 1}$  we may replace 
$\hat{\Gg}_{\lam}$ in \refeq(6-16) 
by $\hat{\Gg}_{\lam}(x)- 
{N^{-1}} \big(\sum_{l=1}^N {\Gg}_{\lam}(x-y_l)\big)
{\bf 1}$ 
without changing the result. However, this changes $e^{-iy_j \xi}$ in 
\refeq(6-16a) to 
\bqn \lbeq(17-1)
\frac1{N}\sum_{l=1}^N (e^{-iy_j \xi}-e^{-iy_l \xi}) 
=
\frac{-i}{N}\sum_{l=1}^N 
\int_0^1 e^{-i\xi(\th{y_j}+ (1-\th){y_l})}d\th \cdot (y_j-y_l)\xi.
\eqn 
Then, \refeq(6-16a) becomes the sum over $j,k,l=1, \dots, N $ of 
\bqn 
U_{jk} \int_{\R^6}
\chi_{\leq {2\ep}}(\xi)\r_{jk}(|\eta|) 
\frac{(y_j-y_l)\cdot \xi}{(\xi^2-\eta^2-i0)|\eta|} 
e^{ix \xi}\la y_k, \hat{\eta}\ra \hat{u}(\eta)d\eta d\xi  \lbeq(16-6b)
\eqn 
where ${\displaystyle U_{jk}=-iN^{-1} \int_0^1 \t_{\th{y_j}+ (1-\th){y_l}} d\th}$
is obviously a good operator. Since 
$(y_j-y_l)\xi= 
(y_j-y_l)\hat{\xi}(|\eta|+ (|\xi|-|\eta|))$, $\hat{\xi}= \xi/|\xi|$, 
we have 
\bqn 
\frac{(y_j-y_l)\xi}{\xi^2-\eta^2-i0}
= (y_j-y_l)\,\hat{\xi}\cdot
\frac{|\eta|}{\xi^2-\eta^2-i0}+ (y_j-y_l)\hat{\xi} \cdot 
\frac{1}{|\xi|+ |\eta|},  \lbeq(16-6c)
\eqn 
which we use in \refeq(16-6b). Then the first term yields  
\bqn  \lbeq(16-7)
\la y_j-y_l, R_x \ra  
\iint_{\R^4} \frac{e^{ix \xi}\chi_{\leq {2\ep}}(\xi)}{(\xi^2-\eta^2-i0)} 
\rho_{jk}(|\eta|)\la y_k, \hat{\eta}\ra \hat{u}(\eta) d\eta d\xi 
\eqn 
If we integrating with respect $\xi$ first and use the polar coordinate 
$\eta= \lam \w$, (16-7) becomes 
\begin{multline}
\la y_j-y_l, R_x \ra \chi_{\leq {2\ep}}(|D|)
\int_{0}^\infty \lam \Gg_{\lam}(x) \left(\int_{\mathbb S^1} 
\Fg(\rho_{jk}(|D|)\la y_k, R\ra u )(\lam\w)d\w \right) d\lam \\
= 2\pi \la y_j-y_l, R_x \ra \chi_{\leq {2\ep}}(|D|) 
K \circ (\rho_{jk}(|D|)\la y_k, R_y \ra) u(x).
\end{multline}
This is a good operator since $\r_{jk}$ is a good multiplier. 
The operator produced by the second term may be expressed as follows:  
\[
\la y_j-y_l, R_x \ra \int_{\R^2}
\left(\iint_{\R^4}
e^{ix\xi-iy\eta}\frac
{\chi_{\leq {2\ep}}(\xi)\chi_{\leq\ep}(|\eta|)}{(|\xi|+ |\eta|)|\eta|g(|\eta|)}d\eta d\xi\right) 
(\tilde{\rho}_{jk}(D)u)(y) dy  
\]
where $\tilde{\rho}_{jk}(|D|)= \tilde{B}_{jk}(|D|)\la y_k, R\ra$ is 
a good operator. 
Thus, the proof of \reflm(2-final) will be completed if we have 
proven the following lemma.

\bglm \lblm(L-integral kernel)
Let $L(x,y)$ be defined by 
\bqn \lbeq(L=def)
L(x,y)= \iint_{\R^4}
e^{ix\xi-iy\eta}\frac
{\chi_{\leq {2\ep}}(\xi)\chi_{\leq\ep}(|\eta|)}{(|\xi|+ |\eta|)|\eta|g(|\eta|)}d\eta d\xi.
\eqn 
Then, the integral operator 
\bqn \lbeq(L-intop)
Lu(x)  = \int_{\R^2}L(x, y) u(y) dy 
\eqn 
is bounded in $L^p(\R^2)$ for $1<p<2$. 
\edlm 
\bgpf The Fourier transform of 
$\chi_{\leq {2\ep}}(\xi)\chi_{\leq\ep}(|\eta|)(|\xi|+ |\eta|)^{-1}$ 
is smooth and in virtue of Lemma B of appendix 
bounded by $C\ax^{-1}\ay^{-1}(\ax+\ay)^{-1}$. 
It follows by virtue of \reflm(plogp) that 
\[
L(x,y) \absleq C \int_{\R^2}\frac{C}
{\ax\la {y-y'}\ra (\ax+ \la {y-y'}\ra)\la y'\ra \log (1+ \la y'\ra)}dy' 
\]
Then, by using Minkowski's inequality twice we have for $1<p<2$ that 
\begin{align} 
& \Big\| \int_{\R^2} L(x,y) |f(y)|dy \Big\|_p 
\leq \int_{\R^2}
\left(\int_{\R^2}|L(x,y)|^p dx \right)^{1/p} |f(y)|dy \notag \\
& \leq C\iint_{\R^4}
\left(\int_{\R^2}
\frac{dx}{\ax^{p}(\ax+ \la {y-y'}\ra)^{p}}\right)^{1/p}
\frac{|f(y)|dy'dy}{\la y'\ra \log (1+ \la y'\ra)\la y-y'\ra} \notag \\
& \leq C \int_{\R^2}
\left(\int_{\R^2}
\frac{dy'}
{\la y-y'\ra^{3-2/p} 
\la y'\ra \log (1+ \la y'\ra)} \right)  
|f(y)|dy   \lbeq(MF)
\end{align} 
We show for the dual exponent $q=p/(p-1)$ of $1< p< 2$, 
\bqn \lbeq(MLQ)
Q(y) \stackrel{\rm def}{=}
\int_{\R^2}
\frac{dy'}
{\la y-y'\ra^{3-2/p} 
\la y'\ra \log (1+ \la y'\ra)} \in L^q(\R^2),  
\eqn
which will prove that the right of \refeq(MF) is bounded by 
$\|Q\|_q \|f\|_p$ by H\"older's inequality and which will complete the proof. 
Since $p>1$, $3-2/p>1$, Schwarz's inequality implies 
\[
|Q(y)|\leq \|\la y\ra^{3-2/p}\|_2 
\|(\la y'\ra \log (1+ \la y'\ra))^{-1}\|_2 < \infty 
\]
and $Q(y)\in L^\infty(\R^2)$. Thus, we need estimate $Q(y)$ 
for large $|y|>100$ only, which we assume in what follows.  
We split $\R^2$ into three regions 
$D_1=\{y' \colon |y'-y|\leq |y|/2\}$, 
$D_2=\{y' \colon  |y-y'|>|y|/2 \ \mbox{and} \  |y'|\leq 2|y|\}$ and 
$D_3=\{y'\colon |y-y'|>|y|/2 \ \mbox{and} \ |y'|>2|y|\}$ 
so that 
\[
Q(y)= 
\left(\int_{D_1} + 
\int_{D_2} +  
\int_{D_3} \right)  
\frac{\la y-y'\ra^{2/p-3} dy'} {\la y'\ra \log (1+ \la y'\ra)} 
\stackrel{\rm def}{=} A(y) + B(y) + C(y). 
\]
On $D_1$ we have $|y|/2<|y'|<2|y|$ and $-1<2/p-2<0$ if 
$1<p< 2$. Hence  
\begin{align*}
A(y) & \leq \frac{C}{\la y\ra \log (1+ \la y\ra)}\int_0^{|y|/2} 
\la r \ra^{2/p-3} r dr 
\leq \frac{C\la y\ra^{2/p-2}}{\log (1+ \la y \ra)}\in L^q(\R^2)  
\end{align*}
since $q(2/p-2)=-2$ and $q>2$. For $B(y)$ we have  
\begin{align*}
B(y)\leq \frac{C}{\la y \ra^{3-\frac2{p}}} \int_0^{2|y|}\frac{rdr}
{\la r\ra \log (2+ r)}
\leq \frac{C}{\la y\ra^{3-\frac2{p}}}
\left(C + \int_{e^2}^{2|y|} \frac{dr}{\log r}\right) 
\end{align*}
Here integration by parts shows 
\[
\int_{e^2}^{2|y|} \frac{dr}{\log r} 
= \left. \frac{r}{\log r}\right]^{2|y|}_{e^{2}} 
+ \int_{e^2}^{2|y|} \frac{dr}{(\log r)^2} 
\leq \frac{2|y|}{\log 2|y|} + \frac12
\int_{e^2}^{2|y|} \frac{dr}{\log r}
\]
and the integral is bounded by $4|y|(\log 2|y|)^{-1}$. It follows once 
more that 
\[
B(y) \leq \frac{C}{\la y\ra^{2-\frac2{p}}\log (1+ \la y \ra)} \in L^q(\R^2)
\]
Finally as $|y'|>2|y|$ implies $|y-y'|>|y'|/2$ and $3-\frac2{p}>1$
\[
C(y) \leq \int_{2|y|}^\infty 
\frac{rdr} {
\la r\ra^{4-\frac{2}{p}} 
\log (1+ \la r \ra)} 
\leq \frac{C}{\la y \ra^{2-\frac{2}{p}}\log (1+ \la y\ra)}
\in L^q(\R^2).
\]
Thus, $Q \in L^q(\R^2)$ as desired and the lemma is proved. 
\edpf 

\subsection{Proof of statement (4)}
We use the notation of \reflm(Step-4). 
By virtue of \reflm(Step-4) $\Ga(\lam)^{-1}$ 
under the assumption of statement (4) satisfies  
\begin{multline}  \lbeq(E-1)
- N^{-1}\lam^{-2} T_e  (T_eG T_e)^{-1} T_e 
-N^{-1}g^{-1}\lam^{-2}T_p (T_e^\perp - \Bg^\ast)T_e^\perp D_0 
T_e^\perp (T_e^\perp - \Bg)T_p \\
+ g^3 T_p M T_p + g^2 S M S + 
g S M + g M S + M.
\end{multline} 
We substitute \refeq(E-1) for $\tGa(\lam)$ of \refeq(low-W), 
which produces seven operators. The one produced by $M(\lam)$ 
is a good operator by the proof of statement (1); those 
produced by $g S M$, $g^2 S M S$ and $g^3 T_p M T_p$ which have 
the factor $S$ {\it on the left} are also good operators. 
This can be seen by repeating the proof statement (2) 
by observing that (i) out of two T's in $TBT$ of \refeq(low-Wa) 
the one {\it on the left} is used for introducing $S$ 
in front of $\hat{I}(\lam)$, which produces \refeq(UseR) with the   
extra factor $\lam$ and that (ii) 
$\lam g(\lam)^j \chi_{\leq \ep}(\lam) M(\lam)$ 
is a good multiplier for any $j\in {\mathbb N}$ and it can play the role 
played by $m(\lam)$ of \refeq(m-def).  

For the operator produced by $ g M(\lam) S$ we have the following lemma. 
\bglm \lblm(rs)
The operator $\W_{rs}$ defined by \refeq(low-Wa) with 
$ g(\lam)M(\lam) S $ in place of $\tGa(\lam)$ is a good operator. 
\edlm
\bgpf Denote by $M(\lam)^\ast$ the conjugate of $M(\lam)$. It suffices 
to prove the lemma when $M(\lam)= 1$ since 
\begin{align}
& \W_{rs}u(x)= 
\frac1{\pi{i}}\int_0^\infty \chi_{\leq\ep} g \lam \big\la 
M(\lam)S\hat{\Gg}_{\lam}(x), 
\int_{\R^2}(\hat{\Gg}_{\lam}(y)-\hat{\Gg}_{-\lam}(y))u(y) dy \big\ra_{\C^N} 
d\lam  \notag  \\ 
& = 
\frac1{\pi{i}}\int_0^\infty \chi_{\leq\ep} g \lam \big\la 
S\hat{\Gg}_{\lam}(x), 
\int_{\R^2}(\hat{\Gg}_{\lam}(y)-\hat{\Gg}_{-\lam}(y))M^\ast(|D|)u(y) 
dy \big\ra_{\C^N} 
d\lam 
\lbeq(rs) 
\end{align}
and $M^\ast(|D|)$ is a good operator. We split $\W_{rs}u$ as follows: 
\[ 
\W_{rs}u = \chi_{\geq 2\ep}(D_x)\W_{rs}u+ \chi_{\leq 2\ep}(D_x)\W_{rs}u.
\] 
(1) We first prove that $\chi_{\geq 2\ep}(D_x)\W_{rs}$ is a good operator. 
For this, it suffices to show the same for the operator $Zu(x)$ defined by  
\bqn \lbeq(Wrs)
Zu(x)= \int_0^\infty \chi_{\leq \ep} \lam g 
\chi_{\geq 2\ep}(|D|){\Gg}_{\lam}(x) 
\left(\int_{\mathbb S}\hat{u}(\lam\w)d\w \right) d\lam  
\eqn 
is a good operato. We repeat the argument of the proof of \reflm(Bad-B). 
Substitute \refeq(chigeq) for $\chi_{\geq 2\ep}(|D|){\Gg}_{\lam}(x)$ in 
\refeq(Wrs), 
which produces two integrals. 
The one produced by $\hat{\m}(x)$ is equal to  
\bqn \lbeq(Wrs-1)
\hat{\mu}(x) 
\int_0^\infty \chi_{\leq \ep}(\lam) \lam g(\lam)
\left(\int_{\mathbb S}(\Fg u)(\lam\w)d\w \right) d\lam 
= \hat{\mu}(x) \la \Fg(\chi_{\ep} {g}), {u}\ra. 
\eqn 
As previously $\hat{\mu} \in L^p(\R^2)$ for all $1\leq p <\infty$ and  
\bqn \lbeq(Fchig)
|\Fg(\chi_{\ep}(|\xi|)g(|\xi|))(x)| 
= |(2\pi)^{-1}(\hat{g}\ast \hat{\chi})(x)| \leq C \ax^{-2}.
\eqn 
Hence, \refeq(Wrs-1) is a good operator. 
Define $\tilde{\m}(\lam)= \lam^2 \chi_{\leq \ep}(\lam)$. Then 
$\tilde{\m}$ is a good multiplier and 
the operator produced by $\m(|D|) \lam^2 {\Gg}_{\lam}(x)$ of 
\refeq(chigeq) is equal to  
\[
{\m(|D|)}\int_0^\infty \chi_{\leq\ep}(\lam) \lam {\Gg}_{\lam}(x) 
\left(
\int_{\mathbb S}\Fg(\tilde{\m}(|D|)u)(\lam\w) d\w \right) d\lam 
=\m(|D|) K \tilde{\m}(|D|)u,  %
\]
which is a good operator. 
Thus, $\chi_{\geq 2\ep}(D_x)\W_{rs}$ is a good operator. \\[5pt]
(2) We next show $\chi_{\leq 2\ep}(D_x)\W_{rs}$ is also a good operator. 
The proof below resembles the one of \reflm(2-final).  
Recall $\t_Y u(x)$ of \refeq(notat-hat). We have  
\begin{align}\lbeq(Wrs-2)
& \chi_{\leq 2\ep}(D_x)\W_{rs}u(x) \\ 
& =\int_0^\infty \chi_{\leq \ep}(\lam) \lam g(\lam)
\Big\la \chi_{\leq 2\ep}(|D|)S \hat{\Gg}_{\lam}(x) 
\int_{\mathbb S}\Fg(M(|D|)\tau_Y u)(\lam\w)d\w \Big\ra_{\C^N}
d\lam.  \notag  
\end{align} 
Here $S \chi_{\leq 2\ep}(|D|)\hat{\Gg}_{\lam}(x)$ is a vector  
whose $j$-th component is given by 
\bqn \lbeq(Kj-def)
K_j(x)= \frac1{N}
\sum_{k=1}^N \frac1{2\pi}\int_{\R^2}
\frac{\chi_{\leq 2\ep}(\xi)e^{ix\xi}(e^{-iy_j\xi}-e^{-iy_k \xi})}
{\xi^2 -\lam^2 -i0} d\xi .
\eqn 
Using Taylor's formula, write 
$e^{-iy_j \xi}-e^{-iy_k\xi}$ in the form 
\[
 -i(y_j-y_k)\xi + |\xi|^2 
\iint_{[0,1]^2} e^{-i\m(\th y_j + (1-\th)y_k)}
(y_j-y_k, \hat{\xi})(\th{y_j}+ (1-\th)y_k, \hat{\xi})
d\th{d\m} 
\]
and define two functions $L_{\lam}^{(1)}(x)$ and $L_{\lam}^{(2)}(x)$ by 
\[
L_{\lam}^{(1)}(x) = \frac1{2\pi}\int_{\R^2}
\frac{\chi_{\leq 2\ep}(\xi)e^{ix\xi}|\xi|}{\xi^2 -\lam^2 -i0} d\xi, \quad 
L_{\lam}^{(2)}(x) = \frac1{2\pi}\int_{\R^2}
\frac{\chi_{\leq 2\ep}(\xi)e^{ix\xi}|\xi|^2}{\xi^2 -\lam^2 -i0} d\xi .
\]
Then, $K_j(x)$ may be expressed as a sum  
\bqn  
K_j(x) = \frac{1}{N}
\sum_{k=1}^N ( -i \la y_j-y_k, R_x\ra L_{\lam}^{(1)}(x)+ 
U_{jk} L_{\lam}^{(2)}(x)). 
\lbeq(Kj1) 
\eqn 
Here $R_x= (R_{x_1}, R_{x_2})$ is the Riesz transform 
and for $j,k=1, \dots, N$  
\bqn 
\tilde{U}_{kj}=  \int_0^1 \int_0^1 \tau_{\m(\th y_k + (1-\th)y_j)}
(y_k-y_j, R_x)(\th{y_k}+ (1-\th)y_j, R_x) d\m {d\th}.
\eqn 
Note that $\la y_j-y_k, R_x\ra$, $U_{kj}$, $M(|D|)$ and 
$\tau_Y$ are all good operators. We shall prove that $Q_1$ and $Q_2$ defined by 
\bqn 
Q_j u(x) = \int_0^\infty \chi_{\leq\ep}(\lam) \lam g(\lam)
L_\lam^{(j)} (x) \left(\int_{\mathbb S}\hat{u}(\lam\w)d\w \right) d\lam, 
\quad j=1,2 
\eqn 
are good operators, which will finish the proof of the lemma. 

As in \refeq(16-6c) we have the identity: 
\[
L_{\lam}^{(1)}(x) = \lam \chi_{\leq 2\ep}(D)\Gg_{\lam}(x) 
+ \frac1{2\pi}\int_{\R^2}
\frac{\chi_{\leq 2\ep}(\xi)e^{ix\xi}}{|\xi|+\lam} d\xi.
\]
Define $\n(\lam) = \lam \chi_{\leq \ep}(\lam)g(\lam)$. Then 
$\n(\lam)$ is a good multiplier and  we have 
\[
Q_1u(x) = \chi_{\leq 2\ep}(D) (K\circ \n(|D|))u(x) + \tilde{L}u(x), 
\]
where $\tilde{L}$ is the integral operator with the integral kernel 
\[
\tilde{L}(x,y) = 
\iint_{\R^4}
e^{ix\xi-iy\eta}\frac
{\chi_{\leq {2\ep}}(\xi)\chi_{\leq\ep}(|\eta|)g(|\eta|)}{4\pi^2 (|\xi|+ |\eta|)}d\eta d\xi.  
\]
It is evident that $\chi_{\leq 2\ep}(D) K\circ (\n(|D|)u(x)$ is a good operator 
and the proof of \reflm(L-integral kernel) implies $\tilde{L}$ is a also 
good operator. Indeed, by using Lemma B in appendix and \refeq(Fchig) we obtain 
\[
|\tilde{L}(x,y)|\leq C \int_{\R^2} 
\frac{dy'}{\ax \la y-y'\ra (\ax+ \la y-y'\ra) \la y'\ra^{2}}
\]
and, the argument which led to \refeq(MLQ) implies  
\[  
\|\tilde{L}f\|_p \leq 
C \int_{\R^2}
\left(\int_{\R^2}
\frac{dy'}
{\la y-y'\ra^{3-2/p}\la y'\ra^2} \right)  
|f(y)|dy.   
\]
Then, Young's inequality implies that the function 
in the parentheses is in $L^q(\R^2)$ for all $1<q<\infty$. 
Thus, H\"older's inequality implies $\tilde{L}$ and, hence 
that $Q_1$ is a good operator. 

We have  
\[
L_\lam^{(2)} (x) = \widehat{\chi_{\leq 2\ep}}(x)+ \lam^2 \Gg_{\lam}(x). 
\]
Define 
$\m(\xi)= \chi_{\leq 2\ep}(|\xi|)g(|\xi|)$ and 
$\tilde\m(\xi)= |\xi|^2 \m(\xi)$. Then $Q_2 u(x)$ is expressed in the form 
\[
Q_2 u(x) = \widehat{\chi_{\leq 2\ep}}(x) 
\left(\int_{\R^2}\la \m(|\xi|)\hat{u}(\xi) d\xi\right) 
+ K\circ \tilde{\mu}(|D|)u, 
\]
$\hat{\mu}\in L^p(\R^2)$ for any $1<p<\infty$ 
and $\tilde{\mu}$ is a good multiplier. 
Thus, $Q_2$ is also a good operator and the proof of \reflm(rs) 
is completed. 
\edpf 

\bglm \lblm(We)
Let $\W_{e}$ be defined by \refeq(low-W) with 
$- N^{-1}\lam^{-2} T_e(T_eG_2 T_e)^{-1} T_e $ 
in place of $\tGa(\lam)$. Then, $\W_e$ is a good operator. 
\edlm 
\bgpf Write 
$H(\lam) = (T_eG_2 T_e)^{-1}$ for shortening formulas . 
Let $\{\eb_1, \dots, \eb_n\}$ be an orthonormal basis of $T_e \C^N$, 
$n = {\rm rank}\, T_e$ and $H_{lm}= \la \eb_l, H \eb_m\ra$, 
$l,m =1, \dots, n$. Then, $W_{e} u(x)$ is equal to 
\begin{align}
& \frac{-1}{N}\int_0^\infty \chi_{\leq\ep}(\lam) \lam^{-1}\big\la 
H(\lam)T_e \hat{\Gg}_{\lam}(x), \int_{{\mathbb S}^1} T_e \widehat
{\t_Y u}(\lam\w) d\w \big\ra_{\C^N} d\lam  \notag \\
& = 
\sum_{lm}\frac{-1}{N}
\int_0^\infty \chi_{\leq\ep} \lam^{-1}H_{lm}
\la \eb_m, \hat{\Gg}_{\lam}(x)\ra 
\left(\int_{{\mathbb S}^1}
\la \eb_l, \widehat
{\t_Y u}(\lam\w) \ra_{\C^N}d\w\right) d\lam 
\lbeq(We)
\end{align} 
Since $S\eb_l= \eb_l$, we may replace in \refeq(We) 
$\widehat{\t_Y u}(\lam\w)$ by 
$S\widehat{\t_Y u}(\lam\w)$ whose $j$-th component 
is given by ${N}^{-1}\sum_{k=1}^N (e^{iy_j {\lam\w}}-e^{iy_k {\lam\w}}) \hat{u}({\lam\w})$. 
As previously we have   
\begin{gather*}
e^{iy_j {\lam\w}}-e^{iy_k {\lam\w}}=  
-i(y_j-y_k){\lam\w} + \lam^2 v_{jk}({{\lam\w}}), \\
v_{jk}({{\lam\w}})= \iint_{[0,1]^2} e^{-i\m(\th y_j + (1-\th)y_k){\lam\w}}
(y_j-y_k, \w)(\th{y_j}+ (1-\th)y_k, \w)
d\th{d\m}. 
\end{gather*}
Since $\eb_l\in T_e \C^N$, we have $\sum_{j=1}^N \eb_l^{(j)}=0$ and 
\[
\la \Gg_1(Y)\eb_l, \eb_l \ra= -\frac{1}{4N}\sum_{j,k=1}^N 
|y_j-y_k|^2 \eb_l^{(j)}\eb_j^{(k)} 
= \frac{1}{2N}\left(\sum_{j=1}^N \eb_l^{(j)}y_j \right)^2=0. 
\]
It follows that $\sum_{k,j=1}^N \eb_l^{(j)}(y_j-y_k){\lam\w} =0$ 
and  
\bqn 
\la \eb_l, \widehat{\t_Y u}(\lam\w) \ra= 
\lam^2 \sum_{k,j=1}^N \eb_l^{(j)} v_{jk}(\lam\w) \hat{u}(\lam\w) 
= \lam^2 \Fg(\Vg_l u)(\lam\w) 
\eqn 
where $\Vg_l $ is a good operator defined by 
\[
\Vg_l u (x) = \sum_{j,k=1}^N \eb_l^{(j)} 
\iint_{[0,1]^2} \t_{\m(\th y_j + (1-\th)y_k)}
(y_j-y_k, R)(\th{y_j}+ (1-\th)y_k, R) u(x). 
\]
Thus, $W_{e} u(x)$ is equal to 
\begin{align*}
& \frac{-1}{N}\sum_{lm n}
\int_0^\infty \chi_{\leq\ep} \lam  H_{lm}(\lam) \eb_m^{(n)}\Gg_{\lam}(x-y_n) 
\left(\int_{{\mathbb S}^1} \Fg(\Vg_l u)(\lam\w)
d\w \right) d\lam \\
&\hspace{1cm} = \frac{-1}{N} \sum_{lm n} \eb_m^{(n)} \t_{y_n} 
K \circ (\chi_{\leq\ep}(|D|)H_{lm}(|D|) \Vg_l u)(x) , 
\end{align*}
which is a good operator. 
\edpf 

Next lemma completes the proof of statement (4) of \refth(Main). 
For shortening the formula we write
\[
B_{\ast\ast}= -\frac{1}{iN\pi}
(T_e^\perp - \Bg^\ast)T_e^\perp D_0 T_e^\perp (T_e^\perp - \Bg).
\]
\bglm \lblm(Wu)
Let $\W_{u}$ be defined by \refeq(low-W) with 
$g^{-1}\lam^{-2}T_p B_{\ast\ast}T_p$ 
in place of $\tGa(\lam)$. Then, $\W_{u}$ is bounded 
in $L^p(\R^2)$ for $1<p \leq 2$ but is unbounded 
for $2<p<\infty$. 
\edlm 
\bgpf By the definition $\W_{u}u(x)$ is equal to 
\bqn 
\int_0^\infty \chi_{\leq \ep}(\lam) g^{-1}\lam^{-1} \big\la 
T_p B_{**}T_p \hat{\Gg}_{\lam}(x), 
\int_{\R^2}(\hat{\Gg}_{\lam}(y)-\hat{\Gg}_{-\lam}(y))u(y) dy \big\ra_{\C^N} 
d\lam 
\lbeq(eqWu)
\eqn 
Then, we repeat the argument of Section 4.3 for $\W_{\rm  low}$ of 
\refeq(low-W-vec) replacing $B_{*}$ by $-i{\pi}N B_{**}$ everywhere.  
Then, if we decompose $W_{u}$ into the sum $W_{u,g}+ W_{u,b}$ as in 
\refeq(gb-decomp) by using the identity \refeq(good-bad), then: \\[5pt]
(i) Proofs of \reflm(good) and \reflm(Bad-B) respectively imply 
without any more changes that $W_{u,g}$ is a good operator and that 
$\chi_{\geq 2\ep}(|D|)W_{u, b}$ is bounded in 
$L^p(\R^2)$ for $1<p<2$. \\[5pt]
(ii) The proof of \reflm(2-final) implies that 
$\chi_{\leq 2\ep}(|D|)W_{u, b}$ is bounded in $L^p(\R^2)$ for $1<p<2$. \\[5pt]
(i) and (ii) should be obvious because the only property 
of $B_\ast$ used in the proof of these lemmas is that 
$T_p B_\ast(\lam)T_p$ is a good multiplier which is shared by 
$T_p B_{**}(\lam)T_p$. \\[5pt]   
(iii) We prove that $\chi_{\geq 2\ep}(|D|)W_{u,b}$ is unbounded 
$L^p(\R^2)$ for $2<p<\infty$ by modifying the argument of 
the proof of \reflm(Bad-C) slightly as follows:  

Let $l={\rm rank}\, D_0 $ and take an orthonormal basis 
$\{\eb_1, \dots, \eb_n\}$ of $T_p\C^N$  
such that $\eb_1, \dots, \eb_l$ are eigenvectors of $T_e^\perp G_1 T_e^\perp$ 
with positive eigenvalues 
and $\eb_{l+1}, \dots, \eb_n \in T_e\C^N $.
Then, equation \refeq(Bad-I) is satisfied 
and the argument after \refeq(Bad-I) implies that, 
for $\chi_{\leq 2\ep}(|D|)W_{u,b}$ to be bounded $L^p(\R^2)$ for $2<p<\infty$,  
it must be satisfied that $B_{**}(|\xi|)R_Y(\w)\r(|\xi|)\in L^p(\R^2)$. 
Since $B_{**}(|\xi|)$ has a bounded inverse 
in $T_e^\perp \C^N$ for small $|\xi|$ such that $\rho(|\xi|)\not=0$, 
it must be then that 
$(T_p \ominus T_e) R_Y(\w)\r(|\xi|)\in L^p(\R^2, T_p\C^N)$ or 
$\la \eb_j, R_Y(\w)\r(|\xi|)\ra \in L^p(\R^2)$, $j=1, \dots, l$. 
However, we have shown in the last part of the proof of \reflm(Bad-C) 
that this is impossible for $j=1, \dots, l$. 
\edpf \edpf 

\subsection{Proof of statement (5) of \refthb(Main)} 
Under the condition of statement (5), $T_p=T_e$ and $\Ga(\lam)^{-1}$
satisfies \refeq(5-1), which we substitute in \refeq(low-W). 
Since $T_p S = ST_p= T_p$, the argument of section 4.4 implies that 
the operator produced by 
$g^2 (T_p M(\lam) S + S M(\lam)T_p)+ 
g (T_p M(\lam)+ g M(\lam)T_p) + M(\lam)$ is a good operator. 
An easy modification of the argument in section 4.2 implies 
that $g^3 T_p M(\lam) T_p$ produces a good operator. 
The argument of the proof of \reflm(We) appliesto show that 
$-N^{-1}\lam^{-2}T_e G_2^{-1} T_e$ 
also produces a good operator. We skip the repetitive details. 
This proves statement (5) and completes the proof of \refth(Main).

\subsection*{Appendix }

In this appendix we show the following lemma: \\[7pt] 
{\bf Lemma A.} {\it For any $\ep>0$ there exits a constant $C_\ep>0$ such that }  
\bqn
\lbeq(Appen)
\int_{\R^4}e^{ix\xi-ip y} 
\frac{\chi_{\leq {2\ep}}(\xi)\chi_{\leq\ep}(|p|)}{|\xi|+ |p|} d\xi{dp}
\leq \frac{C_\ep}{(\ax+ \ay)^3}\log\left(\frac{(\ax+ \ay)^2}{\ax\ay}\right).
\eqn 
\bgpf If we use the identity  
\[
\frac1{|\xi|+|p|} = \int_0^\infty e^{-t(|\xi|+ |p|)} dt 
\]
and Fubini's theorem, then the left side of \refeq(Appen) becomes 
\bqn \lbeq(Appen-2)
\int_0^\infty 
\left(\frac1{2\pi}
\int_{\R^2}e^{ix\xi-t|\xi|} \chi_{\leq {2\ep}}(\xi)d\xi\right) 
\left(\frac1{2\pi}
\int_{\R^2}e^{-ip y-t|p|} \chi_{\leq\ep}(|p|){dp} \right)dt. 
\eqn 
The functions inside parentheses  
are convolutions of the Poisson kernel with bump functions 
$\Fg \chi_{\leq 2{\ep}}(x)$ and $\Fg \chi_{\leq {\ep}}(y)$ 
respectively 
(see \cite{Stein}, p. 61). They are bounded by 
$C_1 t(\ax^2+ t^2)^{-3/2}$ and $C_2 t(\ay^2+ t^2)^{-3/2}$ respectively. 
It follows by changing variable $t$ to $\ax^{1/2}\ay^{1/2} t $ that  
\begin{align} \lbeq(A-3)
\refeq(Appen-2) & \leq 
C \int_0^\infty \frac{t^2 dt}{(\ax^2+ t^2)^{3/2}(\ay^2+ t^2)^{3/2}} \notag \\
& = \frac{C}{\ax^{3/2}\ay^{3/2}}
\int_0^\infty \frac{t^2dt}{(t^4 + s^2 t^2 +1)^{3/2}}, \quad 
s^2= {\frac{\ax^2+\ay^2}{\ax\ay}}. 
\end{align}
We estimate the integral in 
the right hand side of  \refeq(A-3) 
by slitting $(0,\infty)$ into intervals into 
$(0,1/s)$, $(1/s,s)$ and $(s,\infty)$ where the denominator is bounded 
from below by $1$, $s^3 t^3$ and $t^6$ respectively. Then 
\[
\int_0^\infty \frac{t^2dt}{(t^4 + s^2 t^2 +1)^{3/2} } 
\leq \int_0^{1/s}{t^2dt}+ 
\int_{1/s}^s \frac{dt}{s^3 t} + 
\int_{s}^\infty \frac{dt}{t^4} 
= \frac{2}{3s^3}(1+3 \log{s}) .
\]
Since $s\geq \sqrt{2}$, the right side may be further estimated 
by $C s^{-3}\log{s^2}$ and $s^2\leq (\ax+\ay)^2/\ax \ay$. 
Combining this with \refeq(A-3), we obtain the lemma. 
\edpf 

For applications in the text we need only the following weaker version  
which trivially follows from Lemma A. \\[7pt]
{\bf Lemma B.} {\it For any $\ep>0$ there exits a constant $C_\ep>0$ such that }  
\bqn \lbeq(Appen-1)
\frac1{(2\pi)^2} \int_{\R^4}e^{ix\xi-ip y} 
\frac{\chi_{\leq {2\ep}}(\xi)\chi_{\leq\ep}(|p|)}{|\xi|+ |p|} d\xi{dp}
\leq \frac{C_\ep}{\ax\ay(\ax+ \ay)}.
\eqn

\end{document}